\documentclass{IEEEtran}

\usepackage[cmex10]{amsmath}
\usepackage{mathrsfs}
\usepackage{cite}
\usepackage{hyperref}
\usepackage[caption=false,font=footnotesize,
	subrefformat=parens]{subfig}
\usepackage{fp}

\usepackage{amssymb,amsthm}
\usepackage{tikz}
\usetikzlibrary{calc}

\newtheorem{theorem}{Theorem}

\newtheorem{proposition}{Proposition}
\newtheorem{lemma}{Lemma}

\theoremstyle{definition}
\newtheorem{definition}{Definition}

\theoremstyle{remark}

\renewcommand{\qed}{\hfill\IEEEQED}

\newcommand{\CcausalTx}{C^\mathrm{causal}_\mathrm{Tx}}
\newcommand{\CcausalTxRx}{C^\mathrm{causal}_\mathrm{TxRx}}
\newcommand{\CnoncausalTxRx}{C^\mathrm{noncausal}_\mathrm{TxRx}}

\newcommand{\Conline}{T^{\mathrm{online}}}
\newcommand{\Coffline}{T^{\mathrm{offline}}}
\newcommand{\T}{\mathscr{T}}
\newcommand{\Gonline}{\mathcal{G}_n^{\mathrm{online}}}
\newcommand{\Goffline}{\mathcal{G}_n^{\mathrm{offline}}}
\newcommand{\CSmith}{C_{\mathrm{Smith}}}

\DeclareMathOperator*{\argmax}{\arg\!\max}

\begin{document}
\allowdisplaybreaks

% paper title
% can use linebreaks \\ within to get better formatting as desired
% Do not put math or special symbols in the title.
\title{Capacity of the Energy Harvesting Channel with a Finite Battery}

% author names and affiliations
% use a multiple column layout for up to three different
% affiliations
%\author{\IEEEauthorblockN{Dor Shaviv, Phan-Minh Nguyen and Ayfer \"{O}zg\"{u}r}
%\IEEEauthorblockA{Department of Electrical Engineering\\
%Stanford University\\
%Email: \{shaviv, npminh, aozgur\}@stanford.edu}}

\author{Dor Shaviv, Phan-Minh Nguyen and Ayfer \"{O}zg\"{u}r
%\thanks{This work was presented in part at the 2015 IEEE International Symposium on Information Theory (ISIT).}
}

% make the title area
\maketitle

% As a general rule, do not put math, special symbols or citations in the abstract
\begin{abstract}
We consider an energy harvesting channel, in which the transmitter is powered by an exogenous stochastic energy harvesting process $E_t$, such that $0\leq E_t\leq\bar{E}$, which can be stored in a battery of finite  size $\bar{B}$. We provide a simple and insightful formula for the approximate capacity of this channel with bounded guarantee on the approximation gap independent of system parameters. This approximate characterization of the capacity identifies two qualitatively different operating regimes for this channel: in the large battery regime, when $\bar{B}\geq \bar{E}$, the capacity is approximately equal to that of an AWGN channel with an average power constraint equal to the average energy harvesting rate, i.e. it depends only on the mean of $E_t$ and is (almost) independent of the distribution of $E_t$ and the exact value of $\bar{B}$. In particular, this suggests that a battery size $\bar{B}\approx\bar{E}$ is approximately sufficient to extract the infinite battery capacity of the system. In the small battery regime, when $\bar{B}<\bar{E}$, we clarify the dependence of the capacity on the distribution of $E_t$ and the value of $\bar{B}$.  

There are three steps to proving this result which can be of interest in their own right: 1) we characterize the capacity of this channel as an $n$-letter mutual information rate under various assumptions on the availability of energy arrival information: causal and noncausal knowledge of the energy arrivals at the transmitter with and without knowledge at the receiver; 2) we characterize the approximately optimal online power control policy that maximizes the long-term average throughput of the system; 3) we show that the information-theoretic capacity of this channel is equal, within a constant gap, to its long-term average throughput. This last result provides a connection between the information- and communication-theoretic formulations of the energy-harvesting communication problem that have been so far studied in isolation.  %the approximation formulation of the problem as a power control problem for energy-harvesting communication, extensively studied in the communication theory literature over the recent years, provides an upper bound on the true information-theoretic capacity of the channel. For example, the offline power control problem provides an upper bound on the information-theoretic capacity with noncausal knowledge of the energy arrivals at the transmitter and the receiver, while the online problem is an upper bound on the capacity with causal information. Perhaps more surprisingly, we also show that given an optimal power control policy there is a natural way to construct an explicit scheme which achieves a rate within a constant gap of the upper bound for any i.i.d. energy harvesting  process and any battery size. This shows that these two different formulations of the energy-harvesting communication problem, so far studied in isolation, are strongly coupled; solving the power control problem is almost sufficient to solve the information-theoretic problem. 
\end{abstract}

\IEEEpeerreviewmaketitle

%---------------------------------------------------%
\section{Introduction}
%---------------------------------------------------%

Energy-harvesting is quickly becoming a game-changing technology for many wireless systems. The promise of self-sustained perpetual operation opens exciting possibilities for a wide range of applications from powering base stations in rural areas with renewable energy sources (ex. wind or sun) %, and hence enabling telecom networks to expand and operate beyond the limits of the power grid, 
to building in-body wireless networks powered by body heat, motion or RF energy transfer. However, energy harvesting also brings a fundamental shift in communication system design principles.  In conventional systems, energy (or power) is a deterministic
quantity continuously available to the transmitter and communication is typically constrained only in terms
of average power. In harvesting systems, energy may not be generated at all times 
and the rate of energy generation can be unpredictable and fluctuate significantly over time. In such
systems, energy that becomes available for information transmission can be modeled as a stochastic rather
than a deterministic process.

\begin{figure}[!t]
\centering
\begin{tikzpicture}
	\node[draw,rectangle] (Tx) at (0,0) {Transmitter};
	\node[draw,circle] (Sum) at (2.5,0) {$+$};
	\node[draw,rectangle] (Rx) at (5,0) {Receiver};
	\node[draw,rectangle] (Battery) at (0,1) {Battery};
	\node at (-1,1) {$\bar{B}$};
	
	\tikzstyle{every path}=[draw,->]
	\path (Battery) -- (Tx);
	\path (0,1.8) -- node[above,pos=0.1] {$E_t$} (Battery);
	\path (Tx) -- node[above] {$X_t$} (Sum);
	\path (Sum) -- node[above] {$Y_t$} (Rx);
	\path (2.5,1) -- node[above,pos=0.1] {$N_t\sim\mathcal{N}(0,1)$} (Sum);
\end{tikzpicture}
\caption{Energy harvesting AWGN channel model.}
\label{fig:channel}
\end{figure}
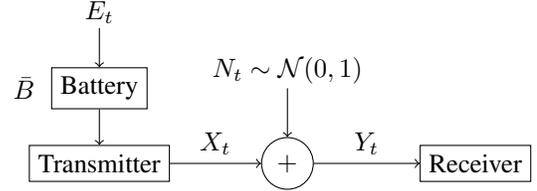

Communication under an average power constraint is well-understood. Shannon's most famous capacity formula
\begin{equation}\label{eq:AWGN}
C= W\log\left(1+\frac{P}{N_0W}\right)\qquad\text{bits/s},
\end{equation}
provides the basis for designing efficient communication systems operating under this constraint. This formula not only quantifies exactly
the performance limit of communication over an additive white gaussian noise (AWGN) channel constrained to an average power of $P$ Watts, allocated bandwidth $W$ Hz, and subject to white noise of power spectral density $N_0/2$ Watts/Hz, but perhaps even more importantly also identifies two fundamentally different operating regimes for this channel where the dependence of the capacity on major system parameters is qualitatively different. In the power-limited (or low-signal-to-noise ratio (SNR)) regime, where $\text{SNR}:=P/N_0W\ll0$ dB, the capacity is approximately linear in
the power, and the performance depends critically on the
 power available but not so much on the bandwidth. In the
bandwidth-limited (or high-SNR) regime, where $\text{SNR}\gg 0$
dB, the capacity is approximately linear in the bandwidth and
the performance depends critically on the bandwidth but not so
much on the power. The regime is determined by the interplay
between the amount of power and bandwidth available.
The design of good communication schemes for wireless systems has been primarily driven
by the parameter regime one is in.

Despite significant recent effort \cite{OzelUlukus2012,MaoHassibi2013,
Tutuncuogluetal2013,DongOzgur2014, JogAnantharam2014,
MaoHassibi2014,Ozeletal2014, DongFarniaOzgur2015}, there is no analogous understanding for energy harvesting  communication systems. Fig.~\ref{fig:channel} depicts the basic model that captures this form of communication. Here a transmitter powered by an exogenous stochastic energy arrival process $E_t$ equipped with a battery of size $\bar{B}$  is communicating to a receiver over an AWGN channel. The available energy for transmission at any given time is limited by the amount of energy available in the battery $B_t$, which in turn depends on the previous energy arrivals as well as the energy consumed in the earlier time slots: 
\begin{align}
	X_t^2&\leq B_t, \nonumber	\\*
	B_t&=\min\{B_{t-1}-X_{t-1}^2+E_t,\bar{B}\}.\label{eq:powerconstEH}
\end{align}
This leads to a complex power constraint on the transmitter which has memory  and is input-dependent. The problem is further complicated by the fact that in a typical scenario the realizations of $E_t$ are known (causally) only at the transmitter and not at the receiver. Obtaining an insightful capacity formula for such a transmitter has proven difficult. As a result, even very basic questions concerning the design of energy harvesting
communication systems remain poorly understood, such as: 
\begin{itemize}
\item how does the system capacity (at least roughly) depend on the energy harvesting profile $E_t$ and the battery size $\bar{B}$? 
\item are there different operating regimes where this dependence is qualitatively different (analogous to the two operating regimes of the classical AWGN channel)? 
\item what are the properties of the process $E_t$ (ex. mean, variance etc.) that most critically determine capacity? as a result, what are more favorable and less favorable energy harvesting profiles? 
\item given an energy harvesting profile $E_t$, how large should we choose the size of  the battery $\bar{B}$ so as to extract most of the system capacity?
\end{itemize}
%The simplest model that captures this form of communication is given in Figure~\ref{fig:channel}. Here a transmitter powered by an exogenous stochastic energy arrival process $E_t$ equipped with a battery of size $\bar{B}$  is communicating to a receiver over an AWGN channel. The energy of the transmitted symbol at each channel use is limited by the available energy in the battery, which is now a random quantity.
Previous work \cite{DongOzgur2014,DongFarniaOzgur2015}  provides an approximate formula for the capacity of this channel when energy arrivals are i.i.d. Bernoulli, which sheds some light on these questions in this specific case. 

\subsection{Our Contribution}
\label{subsec:discussion_of_results}

In this paper, we derive a simple and insightful formula for the approximate capacity of this channel which holds for any i.i.d. process $E_t$. In particular, we show that when $E_t$ is an i.i.d process, the capacity of this channel can be approximated as 
\begin{equation}\label{eq:mainres}
C\approx \frac{1}{2}\log\left(1+\mathbb{E}[\min\{E_t,\bar{B}\}]\right)\qquad\text{bits/s/Hz}
\end{equation}
within $3.85$ bits/s/Hz, where we assume that the noise variance is normalized to $1$.\footnote{Different from \eqref{eq:AWGN}, here we focus on the capacity in bits/s/Hz and assume that the noise variance is normalized to $1$ in order to highlight the dependence of the capacity on the new channel parameters. The normalization of the noise power to $1$ is without loss of generality since the approximation result is independent of the parameters of the problem.}\footnote{The additive capacity approximations we develop in this paper are most relevant in the high-SNR regime. Note that this can indeed be the operating regime of a low power wireless device if the available power is concentrated on a very narrow frequency band. This ultra-narrow band approach is indeed the defacto technique for some low power IoE devices that, despite being low-power, operate at reasonably high SNRs \cite{sigfox, TI}.} This characterization identifies two fundamentally different operating regimes for this channel where the dependence of the capacity on $E_t$ and $\bar{B}$ is qualitatively different. 

Let $E_t$ take values in the interval $[0, \bar{E}]$. When $\bar{B}\geq \bar{E}$, \eqref{eq:mainres}~becomes
\begin{equation}\label{eq:mainres1}
C\approx \frac{1}{2}\log\left(1+\mathbb{E}[E_t]\right)\qquad\text{bits/s/Hz},
\end{equation}
and the capacity is approximately equal to that of an AWGN channel with an average power constraint equal to the average energy harvesting rate. Note that the right-hand side of \eqref{eq:mainres1} trivially upper bounds the capacity of the energy harvesting channel as this would be the capacity if the transmitter were only constrained in its average transmission power (or average energy per channel use), which can obviously not exceed the average rate of the incoming energy. Ozel and Ulukus in \cite{OzelUlukus2012} show that this upper bound can be achieved when $\bar{B}=\infty$. Our result suggests that this large battery regime kicks in much earlier, as soon as the battery size $\bar{B}$ is large enough to accommodate the maximal amount of energy that can be harvested over a single channel use. This is surprising given that the transmitter is limited by the additional constraint \eqref{eq:powerconstEH},  and at finite $\bar{B}$ this can lead to part of the harvested energy being wasted due to an overflow in the battery capacity. While there is a very natural way to achieve the AWGN capacity in \eqref{eq:mainres1} with $\bar{B}=\infty$ -- the transmitter can simply remain silent for a duration of time sublinear in the blocklength to accumulate sufficient energy in the battery and then use standard i.i.d. Gaussian coding -- achieving the AWGN capacity at finite $\bar{B}$ is intricate and in particular requires an optimal online power control strategy. Note that in this large battery regime, which \eqref{eq:mainres} identifies as the case  $\bar{B}\geq \bar{E}$, the capacity approximation depends only on the mean of the energy harvesting process: two energy harvesting profiles are equivalent as long as they provide the same energy on the average. The approximate capacity is also independent of the exact size of $\bar{B}$. In particular, choosing $\bar{B}\approx \bar{E}$ is almost sufficient to extract the infinite battery capacity. More precisely, there is limited capacity gain in making $\bar{B}$ much larger than $\bar{E}$. 

\begin{figure}[!t]
\centering
\subfloat[$\bar{B}\geq\bar{E}$]{
\begin{tikzpicture}
\tikzstyle{every node}=[font=\scriptsize];

% axes
\draw[->] (0,0) -- (3,0);
\draw[->] (0,0) -- (0,2);
\node[below] at (0,0) {$0$};
\node[above] at (0,2) {$f_E(x)$};
\node[right] at (3,0) {$x$};

% distribution
\draw[red,thick] (0,0.75) to[out=0,in=180] (0.75,1.25);
\draw[red,thick] (0.75,1.25) to[out=0,in=135] (1.75,0);
\node[below] at (1.75,-0.1) {$\bar{E}$};
\draw (1.75,0) -- (1.75,-0.1);

% Bmax
\draw[dashed,thick] (2.5,0) -- (2.5,2);
\node[below] at (2.5,-0.1) {$\bar{B}$};
\draw (2.5,0) -- (2.5,-0.1);

\end{tikzpicture}
\label{subfig:large_battery_regime}
}
\ 
\subfloat[$\bar{B}\leq\bar{E}$]{
\begin{tikzpicture}

\tikzstyle{every node}=[font=\scriptsize];

% axes
\draw[->] (0,0) -- (3,0);
\draw[->] (0,0) -- (0,2);
\node[below] at (0,0) {$0$};
\node[above] at (0,2) {$f_E(x)$};
\node[right] at (3,0) {$x$};

% distribution
\draw[red,thick] (0,0.75) to[out=0,in=180] (0.75,1.25);
\draw[red,thick] (0.75,1.25) to[out=0,in=140] (1.1,1.12);

% Bmax
\node[below] at (1.1,-0.1) {$\bar{B}$};
\draw (1.1,-0.1) -- (1.1,0);
\draw[->,red,thick,>=latex] (1.1,0) -- (1.1,1.5);

\end{tikzpicture}
\label{subfig:small_battery_regime}
}
\caption{pdf of $E_t$ in the two battery regimes.}
\label{fig:two_battery_regimes}
\end{figure}
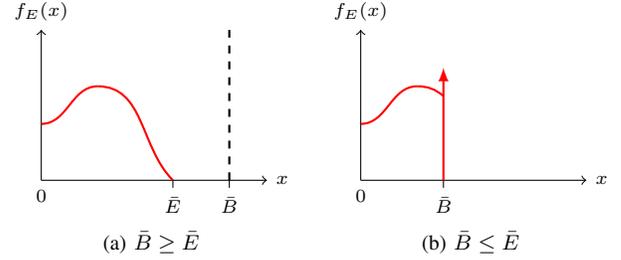

When $\bar{B}\leq \bar{E}$, note that one can equivalently consider the distribution of $E_t$ to be that in Fig.~\ref{fig:two_battery_regimes}-\subref{subfig:small_battery_regime}: since every energy arrival with value $E_t\geq\bar{B}$ fully recharges the battery, this creates a point mass at $\bar{B}$ with value $P(E_t\geq\bar{B})$. In this case, \eqref{eq:mainres} reveals that the capacity is approximately given by the mean of this modified distribution. This can be interpreted as the small battery regime of the channel. In particular, in this regime the capacity roughly depends both on the shape of the distribution of $E_t$ and the value of $\bar{B}$. For example, while two energy harvesting mechanisms providing the same average energy lead to the same approximate capacity in the large battery regime, in the small battery regime they are likely to yield different capacity approximations. In particular, constant energy arrivals will maximize capacity among all distributions with the same mean. Note that while \eqref{eq:mainres} suggests that choosing $B\approx \bar{E}$ allows to almost extract the infinite battery capacity of the channel, it also quantifies the performance loss when $\bar{B}<\bar{E}$. In particular, the performance loss is dictated by the difference between the mean of the original distribution for $E_t$ in Fig.~\ref{fig:two_battery_regimes}-\subref{subfig:large_battery_regime} and that of the modified distribution in Fig.~\ref{fig:two_battery_regimes}-\subref{subfig:small_battery_regime}. Note that  when $E_t$ is unbounded but has  a fast decaying tail, choosing $\bar{B}$ large but finite can be sufficient to approach the infinite battery capacity.

\bigbreak

There are three major steps to proving the approximation result in \eqref{eq:mainres} which can be of interest in their own right. We first characterize the exact capacity of the energy harvesting channel in Fig.~\ref{fig:channel} as the limit of an $n$-letter maximum mutual information rate under various assumptions on the available information regarding the energy harvesting process, such as causal or noncausal information at the transmitter with or without information at the receiver.
Obtaining  an  $n$-letter mutual information expression for the capacity of this channel is nontrivial and has remained an open problem until now, since the energy constraints on the transmitter lead to an input-dependent random state with memory for the system. In particular, it is not a priori clear if the channel is \emph{information-stable} or not. Earlier characterizations of the capacity \cite{MaoHassibi2014} have been only available  in terms of the Verd\'{u}-Han framework~\cite{VerduHan1994}.

The second step is to devise an approximately optimal online power control policy for the energy harvesting channel that maximizes its long-term average throughput. Optimal power control for energy harvesting systems has been of significant interest over the recent years in the communication theory literature \cite{YangUlukus2012,Ozeletal2011,TutuncuogluYener2012,online_infiniteB1,online_infiniteB2,online_infiniteB3,online_infiniteB4,info2013,DP0,DP1,DP2,DP3, Mitran}. This formulation simplifies the communication problem by assuming that there is an
underlying transmission scheme operating at a finer time-scale such that allocating power $P$ to this scheme
yields an information rate $r(P) = \frac{1}{2}\log(1 + P)$ bits/s and focuses on maximizing the long-term average throughput of the system subject to energy availability constraints imposed by the harvesting process analogous to \eqref{eq:powerconstEH}. The problem has been studied in two different settings. In the offline case, where energy arrivals are known ahead of time, the optimal power control policy has been characterized in \cite{YangUlukus2012,Ozeletal2011,TutuncuogluYener2012}. The optimal strategy in this case keeps  energy consumption as constant as possible over time while ensuring  energy is never wasted due to an overflow in the battery capacity. The more interesting online case remains poorly understood. In this setting, the energy arrival process is observed causally at the transmitter and only statistical information regarding the future energy arrivals is available. In this paper, we develop a simple online power control policy based on~\cite{DongOzgur2014,DongFarniaOzgur2015} and show that it is at most 1.8 bits/s/Hz to optimality, independent of system parameters. This strategy waits until the battery is fully charged to $\bar{B}$ and then allocates the energy $\bar{B}$ in an exponentially decaying manner until the next time instant when the battery is fully recharged.
%This solution reveals that the optimal power control strategy in the online case is structurally different from the offline case, where the optimal policy rather aims to keep power allocation as constant as possible across different time-slots.
This near-optimal solution suggests that the optimal power control strategy in the online case is structurally different from the offline case, where the optimal policy rather aims to keep power allocation as constant as possible across different time-slots.

The third step in proving \eqref{eq:mainres} is to connect the two different formulations of the energy harvesting communication problem discussed above: the information-theoretic formulation which aims to characterize the fundamental capacity of the channel, and the communication-theoretic power control problem which aims to maximize the long-term average throughput of the channel. We first show that the long-term average throughput provides an upper bound on the true information-theoretic capacity of the channel. For example, the optimal offline throughput provides an upper bound on the information-theoretic capacity with noncausal knowledge of the energy arrivals at the transmitter  and the receiver, while the optimal online throughput is an upper bound on the capacity with causal  information regarding the energy arrivals. Perhaps more surprisingly, we also show that given an optimal power control policy there is a natural way to construct explicit schemes which achieve a rate within a constant gap of the upper bound (the corresponding long-term average throughput) for any i.i.d. energy harvesting process and any value of $\bar{B}$. This allows us to conclude that the optimal solutions of these two problems, the information-theoretic capacity and the long-term average throughput, cannot differ from each other by a constant gap (ex. the gap is at most $1.05$ bits/s/Hz with receiver side information). In particular, using the approximately optimal online power control policy we develop in the previous step, we obtain the approximation~\eqref{eq:mainres} for the information theoretic capacity.  To the best of our knowledge, this is the first work that establishes an explicit relation between the two different formulations of the energy harvesting communication problem, so far studied separately in the respective information and communication theory literatures.

\subsection{Related Work}
\label{subsec:related_work}

The information-theoretic capacity of the energy harvesting channel in Fig.~\ref{fig:channel} has been of significant recent interest
\cite{OzelUlukus2012,MaoHassibi2013,
Tutuncuogluetal2013,DongOzgur2014, JogAnantharam2014,
MaoHassibi2014,Ozeletal2014, DongFarniaOzgur2015}. In particular, \cite{OzelUlukus2012} shows that when $\bar{B}=\infty$ the capacity of the energy harvesting channel is the same as that of an AWGN channel with average power
constraint equal to the average energy harvesting rate $\mathbb{E}[E_t]$. Follow-up works provide upper and lower bounds on the capacity for the more realistic case of finite battery and {$n$-letter} capacity expressions for some special cases. In particular, \cite{MaoHassibi2013} provides an expression for the capacity in terms of the Verd\'{u}-Han framework; \cite{MaoHassibi2014} derives upper and lower bounds on the capacity with i.i.d. energy arrivals in terms of limits of  $n$-letter maximum mutual information rates; \cite{JogAnantharam2014} considers the special  case when there is a constant amount of energy arriving at each time slot and provides an $n$-letter expression for the capacity in this deterministic case; \cite{Tutuncuogluetal2013} considers the special case where the battery is of unit size and the channel is a noiseless bit pipe. Characterizing the capacity, even as an $n$-letter mutual information rate, has remained an open problem for general energy harvesting processes. The contribution of the current paper with respect to this literature can be regarded as: 1) providing $n$-letter expressions for the capacity under various assumptions on the available information at the transmitter and the receiver regarding the energy arrivals; 2) provide a simple and insightful approximation formula for the capacity of this channel by proving upper and lower bounds that differ by a constant gap. Our work is most closely related to prior work in \cite{DongOzgur2014,DongFarniaOzgur2015} which introduces the constant gap approximation approach and provides an approximate formula for the capacity of the channel in Fig.~\ref{fig:channel} when the energy arrival process is i.i.d. Bernoulli (without providing an ${n\text{-letter}}$ expression for the channel capacity). In a companion paper \cite{ShavivOzgur2015}, we focus on the special case with Bernoulli energy recharges. For this special case, we are able to provide $n$-letter expressions in simpler form, for which the maximizing input distribution can be identified in certain cases, and we exactly solve the corresponding online power control problem. This leads to tighter approximations of the capacity in this special case and allow us to deduce new insights on the usefulness of noncausal observations of the energy arrivals and  output feedback (see also \cite{ShavivOzgurPermuter2015}).

The study of optimal power control policies for the energy harvesting channel precedes the study of its information-theoretic capacity. The power control problem is well understood in the offline case \cite{YangUlukus2012,Ozeletal2011,TutuncuogluYener2012}. The online power control problem can be cast as a Markov Decision Process and the optimal solution can be computed numerically using dynamic programming \cite{DP0,DP1,DP2,DP3}. However, the curse of dimensionality inherent in the dynamic programming solution makes this approach computationally intensive. More importantly, the numerical solution provides little insight into the structure of the optimal power control strategy, its dependence on major system parameters, and the resultant performance. Several works focus on establishing properties for the optimal solution, however these properties are either very high-level, ex. \cite{DP2} establishes monotonicity of the optimal policy, \cite{Wang15} shows that deterministic policies are sufficient;  or still require numerical evaluation, ex. \cite{Mitran} derives a system of coupled partial integro-differential equations as necessary conditions for optimality, which can be solved only numerically. It is easy to observe that when the battery size is infinite, a simple power control strategy that allocates constant power equal to the mean energy arrival rate becomes asymptotically optimal and achieves the AWGN capacity for any i.i.d. energy harvesting process. \cite{online_infiniteB1,online_infiniteB2,online_infiniteB3,online_infiniteB4,info2013} study the infinite battery regime in more detail. Finally, \cite{Ozeletal2011} and follow-up work propose heuristics without providing any guarantees on optimality. In contrast, in this paper we propose an explicit online power control policy and show that it is within a constant gap to optimality. Indeed, this policy can achieve the AWGN capacity, which is achievable with infinite battery size, within a constant gap with finite battery. The structure of this approximately optimal policy is completely different from the heuristics proposed in the literature, which are typically inspired by the offline solution or the infinite battery regime.

%The paper is organized as follows: the channel model is presented in Section~\ref{sec:system_model}.
%Section~\ref{sec:equivalent_channel_and_main_results} contains the main results of the paper. 
%The proof for the capacity expressions is provided in Section~\ref{sec:capacity_proof},
%and Section~\ref{sec:capacity_bounds} provides derivation of upper and lower bounds.
%In Section~\ref{sec:initial_battery_state} we show that the initial state of the battery does not affect the capacity,
%and finally, Section~\ref{sec:Bernoulli_reduction} provides a simple yet important special case of the energy harvesting channel.

\subsection{Organization of the Paper}
The paper is organized as follows. Section~\ref{sec:system_model} contains our system model and Section~\ref{sec:main_results} contains our main results. Section~\ref{sec:capacity_proof} characterizes the capacity of the energy harvesting channel as an $n$-letter mutual information rate. Section~\ref{sec:power_control} considers the online power control problem and develops an approximately optimal power control policy. The reader interested in the power control problem and not the information-theoretic capacity of this channel can read this section independently. Finally Section~\ref{sec:capacity_bounds} develops the connection between the information-theoretic capacity and the power control optimization problem.

%---------------------------%
\section{System Model}
%---------------------------%
\label{sec:system_model}

We begin by introducing the notation used throughout the paper. Let uppercase, lowercase, and calligraphic letters denote random variables, specific realizations of RVs, and alphabets, respectively.
For two jointly distributed RVs $(X,Y)$, let $P_{X}$, $P_{X,Y}$, and $P_{Y|X}$, respectively denote the marginal of $X$, the joint distribution of $(X,Y)$, and the conditional distribution of $Y$ given $X$.
Let $\mathbb{E}[\cdot]$ denote expectation.
For $m\leq n$, $X_m^n=(X_m,X_{m+1},\ldots,X_{n-1},X_n)$, and $X^n=X_1^n$.
With abuse of notation, when the superscript is 2 it is understood as square, i.e. $X_t^2=(X_t)^2$.
Additionally, when the length is clear from the context, we sometimes denote vectors by boldface letters, e.g. $\mathbf{x}\in\mathcal{X}^n$.
All logarithms are to base 2 ($\ln$ will denote log to base $e$).

The energy harvesting channel is an AWGN channel, i.e. the output at time $t$ is $Y_t=X_t+N_t$,
where $N_t\sim\mathcal{N}(0,1)$ and $X_t\in\mathbb{R}$ is the input.
The transmitter has a battery with finite capacity $\bar{B}$,
and the input symbol energy at each time slot is constrained by the available energy in the battery. Let $B_t$ represent the available energy in the battery at time $t$.
The system energy constraints can be described as
\begin{align}
	X_t^2&\leq B_t, 	\label{eq:EH_constraint}\\
	B_t&=\min\{B_{t-1}-X_{t-1}^2+E_t,\bar{B}\}\label{eq:EH_battery}.
\end{align}
$E_t$ is the energy arrivals process which we assume to be i.i.d. for different $t$. For simplicity we assume that $E_t$ is a discrete RV over the finite alphabet $\mathcal{E}$, such that $E_t\geq0$ and $E_t>0$ with positive probability, implying $\mathbb{E}[E_t]>0$. However, our results also apply when $E_t$ is continuous or comes from a mixed distribution by considering the continuous distribution to be the limit of quantized discrete distributions (see footnote~\ref{foot:continuous_E}). We assume that $B_0=b$, where $0\leq b\leq \bar{B}$ is a fixed quantity known to both the transmitter and the receiver before beginning of transmission.\footnote{\label{foot:initial_battery_state}This assumption is made to simplify the exposition. Our results hold even if the initial battery state is unknown to both the transmitter and the receiver and can be arbitrarily fixed or random.}

In this work, we investigate three cases:
$E_t$ is observed causally at the transmitter only;
$E_t$ is observed causally at the transmitter as well as the receiver;
and $E_t$ is observed noncausally at the transmitter and the receiver.
In any case, the transmitter has (at least) causal knowledge of $E_t$, which implies also causal knowledge of $B_t$. While it is natural for the transmitter to be aware of its own energy arrival process in a causal fashion, the receiver may have side information regarding the energy arrival process at the transmitter in certain scenarios; for example, when it is itself harvesting energy from a correlated process or when the transmitter indeed harvests the RF energy dissipated by the receiver. 

For the first case, we define
an $(M,n,\varepsilon)$ code as a set of encoding functions $f^{\mathrm{enc}}_t$ and a decoding function $f^{\mathrm{dec}}$:
\begin{align}
	f^{\mathrm{enc}}_t&:\mathcal{M}\times \mathcal{E}^t\to\mathcal{X}
		,\qquad t=1,\ldots,n,\label{eq:EH_encoding}\\*
	f^{\mathrm{dec}}&:\mathcal{Y}^n\to\mathcal{M},
		\label{eq:EH_decoding}
\end{align}
where $\mathcal{X}=\mathcal{Y}=\mathbb{R}$
and $\mathcal{M}=\{1,\ldots,M\}$.
To transmit message $w\in\mathcal{M}$ at time $t=1,\ldots,n$, the transmitter sets $X_t=f^{\mathrm{enc}}_t(w,E^t)$.
The battery state $B_t$ is a deterministic function of $(X^{t-1},E^t)$, therefore also of $(w,E^t)$.
The functions $f^{\mathrm{enc}}_t$ must satisfy the energy constraint~\eqref{eq:EH_constraint}:
\[(f^{\mathrm{enc}}_t(w,E^t))^2\leq B_t(w,E^t).\]
The receiver sets $\hat{W}=f^{\mathrm{dec}}(Y^n)$.
The probability of error is
$$P_e=\frac{1}{M}\sum_{w=1}^{M}\Pr(\hat{W}\neq w\ |\ w\text{ was transmitted})\leq\varepsilon.$$
%The capacity is defined in the standard way as the supremum of achievable rates.
The rate of an $(M,n,\varepsilon)$ code is $\frac{\log M}{n}$.
We say $R$ is \mbox{$\varepsilon$-achievable} if for every $\delta>0$
there exist, for all sufficiently large $n$, an $(M,n,\varepsilon)$ code with rate $\frac{\log M}{n}>R-\delta$.
The capacity $C$ is the maximal rate that is $\varepsilon$-achievable for all $0<\varepsilon<1$.

When $E_t$ is observed also at the receiver (either causally or noncausally), \eqref{eq:EH_decoding} is altered to
$f^{\mathrm{dec}}:\mathcal{Y}^n\times\mathcal{E}^n\to\mathcal{M}$.
Similarly, to account for noncausal observations of $E_t$ at the transmitter, we change~\eqref{eq:EH_encoding} to
$f^{\mathrm{enc}}:\mathcal{M}\times\mathcal{E}^n\to\mathcal{X}^n$,
so that $X^n=f^{\mathrm{enc}}(w,E^n)$, where again
\[(f^{\mathrm{enc}}_t(w,E^n))^2\leq B_t(w,E^n).\]
Note that $B_t$ is again deterministic function of $(X^{t-1},E^t)$, which is now a deterministic function $(w,E^n)$.

In what follows, we investigate three cases: energy arrival process observed causally at the transmitter only; observed causally at the transmitter as well as the receiver; and observed noncausally at both the transmitter and the receiver. 
We obtain capacities for each of these cases, denoted by $\CcausalTx$, $\CcausalTxRx$, and $\CnoncausalTxRx$, respectively.

%---------------------------%
\subsection{Equivalent Channel Model with Causal TX Side Information}
%---------------------------%
\label{sec:equivalent_channel_and_main_results}
Consider the channel defined in the previous section with the energy arrivals observed causally at the transmitter. Following Shannon's approach~\cite{Shannon1958}
% (see \cite{MaoHassibi2013} for a detailed development),
as done in~\cite{MaoHassibi2013},
this channel can be converted into an equivalent channel with no state information at the transmitter but with a different input alphabet, using \emph{Shannon strategies}: the input to the equivalent channel at time $t$ is a function 
$
u_t:\mathcal{E}^t\to\mathcal{X}
$
and the input alphabet for blocklength $n$ is of the form
\begin{equation}\label{eq:inputalphabet}
	\mathcal{U}^n=\{u^n|\ u_t:\mathcal{E}^t\to\mathcal{X},\ 
	t=1,\ldots,n\}.
\end{equation}
Note that $\mathcal{U}^n$ is not a Cartesian product of $n$ copies of a single alphabet, but a set of $n$-tuples where each element is defined above.
At time $t$, given the realization of $E^t$,  $X_t=U_t(E^t)$ is transmitted over the original channel. The output of the channel is the corresponding $Y_t\in\mathcal{Y}$. This implies the following transition probabilities for this new channel:
\begin{align}
	P_{Y^n|U^n}(y^n|u^n)
	%&=\sum_{e^n}P_{E^n}(e^n)P_{Y^n|U^n,E^n}(y^n|u^n(e^n),e^n)\nonumber\\
	&=\sum_{e^n}P_{E^n}(e^n)P_{Y^n|X^n}(y^n|u^n(e^n))\nonumber\\*
	&=\sum_{e^n}\prod_{t=1}^{n}P_E(e_t)P_{Y|X}(y_t|u_t(e^t)).
	\label{eq:U_transition_prob}
\end{align}
Note that there is no transmitter side information for this channel and the encoding functions~\eqref{eq:EH_encoding} become $f^{\mathrm{enc}}:\mathcal{M}\to\mathcal{U}^n$. However, not all n-tuples in $\mathcal{U}^n$ are admissible. The energy constraints on our original energy harvesting channel imply that the admissible channel inputs $u^n$ should satisfy for every $e^n\in\mathcal{E}^n$:
\begin{align}
	(u_t(e^t))^2&\leq b_t, \label{eq:U_energy} \\ 
	b_t&=\min\{b_{t-1}-(u_{t-1}(e^{t-1}))^2+e_t,\bar{B}\}.
	\label{eq:U_battery}
\end{align}
It is easy to see that the capacity of this channel is equal to that of our original channel, as coding strategies for one can be immediately translated to the other. 

%When the transmitter has causal information of the energy arrivals, we consider the equivalent channel described above. When the transmitter has noncasual information, we consider the original channel model. 
When the energy arrival process is observed only at the transmitter, we consider the equivalent channel described above instead of the original model. When the receiver also observes the energy arrival process, we consider the original channel model. 

\section{Main Results}\label{sec:main_results}

The main result of this paper is the approximation of the capacity of the energy harvesting channel under various assumptions on the availability of energy arrival information at the transmitter and the receiver given in the following theorem.

\begin{theorem}
\label{thm:main_theorem}
The capacity of the energy harvesting channel in bits/channel use is bounded by
\begin{align}
\frac{1}{2}\log(1+\mu)- 3.85 &\leq \CcausalTx
\leq \frac{1}{2}\log(1+\mu),
\label{eq:C_Tx_approx}\\
\frac{1}{2}\log(1+\mu)-2.85&\leq \CcausalTxRx,\CnoncausalTxRx
\leq \frac{1}{2}\log(1+\mu),
\label{eq:C_TxRx_approx}
\end{align}
where $\mu\triangleq \mathbb{E}[\min\{E_t,\bar{B}\}]$.
\end{theorem}

The proof of this theorem consists of three main steps, each of which can be of interest in its own right. The first step is to characterize the capacity of the energy harvesting channel as an $n$-letter mutual information rate under various assumptions on the availability of energy arrival information (Theorem~\ref{thm:capacity}). The second step is to use this characterization to show that the information-theoretic capacity of the energy harvesting channel is within a constant gap of 1.05 or 2.05 (respectively for \eqref{eq:C_Tx_approx} and \eqref{eq:C_TxRx_approx}) from its power control formulation studied in the communication theory literature (Theorem~\ref{thm:bounds}).  The third step is to provide an approximate solution to the online power control problem with bounded guarantee of 1.8 on the approximation gap (Theorem~\ref{thm:onlinePC}).

To state the expressions for capacity, we define the set of allowed input distributions on $\mathcal{U}^n$ for the equivalent channel:
\begin{align}
	\mathcal{P}_n(b)=\Big\{&P_{U^n}
		\text{ s.t. a.s. for $t=1,\ldots,n$ and 
		$\forall e^n\in\mathcal{E}^n$}:\nonumber\\*
		&(U_t(e^t))^2\leq B_t,\ B_0=b, \nonumber\\*
		&B_t=\min\{B_{t-1}-(U_{t-1}(e^{t-1}))^2+e_t,\bar{B}\}\Big\}.
		\label{eq:U_constraint}
\end{align}
Note that we impose the energy constraints by assigning zero probability to any codeword that does not satisfy~\eqref{eq:U_energy} and~\eqref{eq:U_battery}. Similarly, define 
\begin{align}
	\mathcal{F}_n(b)=\Big\{&P_{X^n|E^n}
	\text{ s.t. }\forall e^n\in\mathcal{E}^n,
	\text{ a.s. for }t=1,\ldots,n:\nonumber\\*
	&X_t^2\leq B_t,\ B_0=b,\nonumber\\*
	&B_t=\min\{B_{t-1}-X_{t-1}^2+e_t,
		\bar{B}\} \Big\}.
	\label{eq:noncausal_Fn_definition}
\end{align}
%This is similar to the definition in~\cite{Ozeletal2014}, but with the initial state of the battery given as a parameter.

For the case of causal energy arrival information at both the transmitter and the receiver, we use the notion of causal conditioning as in~\cite{Kramer2003}, namely let
\begin{equation}
\label{eq:causal_conditioning}
	P_{X^n\|E^n}(x^n\|e^n)
	\triangleq\prod_{t=1}^{n}P_{X_t|X^{t-1},E^t}
	(x_t|x^{t-1},e^t).
\end{equation}
This differs from $P_{X^n|E^n}=\prod_{t=1}^{n}P_{X_t|X^{t-1},E^n}(x_t|x^{t-1},e^n)$ in that at time $t$ the dependence on $E^n$ is replaced by only the past and present $E^t$.
%This is more restrictive than $P_{X^n|E^n}$, in the sense that $X_t$ is independent of the future energy arrivals $E_{t+1}^{n}$ given $(X^{t-1},E^t)$.
Define
\begin{align}
	\mathcal{Q}_n(b)=\Big\{P_{X^n\|E^n}:\ 
	&P_{X^n|E^n}=P_{X^n\|E^n}
	\text{ s.t. }\nonumber\\*
	&\forall e^n\in\mathcal{E}^n
	\text{ a.s. for }t=1,\ldots,n:\nonumber\\*
	&X_t^2\leq B_t,\ B_0=b,\nonumber\\*
	&B_t=\min\{B_{t-1}-X_{t-1}^2+e_t,
		\bar{B}\} \Big\}.
	\label{eq:Qn_definition}
\end{align}
This is similar to $\mathcal{F}_n(b)$, but imposes the additional constraint that $X_t$ must depend on $E_t$ in a causal manner, as defined in~\eqref{eq:causal_conditioning}.
Note that $B_t$ is a function of $(X^{t-1},E^t)$, so $\mathcal{Q}_n(b)$ is well-defined.

Using these definitions, we state the following theorem:
\begin{theorem}\label{thm:capacity}
The capacities of the energy harvesting channel with various levels of energy arrival information are given by
\begin{align}
	\CcausalTx&=\lim_{n\to\infty}\frac{1}{n}
		\sup_{P_{U^n}\in\mathcal{P}_n(b)}I(U^n;Y^n),
		\label{eq:causal_capacity}\\
	\CcausalTxRx&=\lim_{n\to\infty}\frac{1}{n}
		\sup_{P_{X^n\|E^{n}}\in\mathcal{Q}_n(b)}
		I(X^n;Y^n|E^n),
		\label{eq:causalRx_capacity}\\
	\CnoncausalTxRx&=\lim_{n\to\infty}
		\frac{1}{n}\sup_{P_{X^n|E^n}\in\mathcal{F}_n(b)}
		I(X^n;Y^n|E^n),
		\label{eq:noncausal_capacity}
\end{align}
where the supremum in~\eqref{eq:causalRx_capacity} should be interpreted as setting the input distribution $P_{X^n|E^n}(x^n|e^n)=P_{X^n\|E^n}(x^n\|e^n)$, or in other words, the Markov chain $X_t-(X^{t-1},E^{t-1})-E_{t+1}^n$ holds for every $t=1,\ldots,n$.
\end{theorem}

Although we focus on the AWGN channel in this paper, it is straightforward to see that Theorem~\ref{thm:capacity} generalizes to any memoryless channel. The proof of the theorem is given in Section~\ref{sec:capacity_proof}.

The expressions in Theorem~\ref{thm:capacity} depend on the initial state of the battery $B_0=b$.
However, in the following, we show that the capacity does not depend on~$b$, which implies that the expressions~\eqref{eq:causal_capacity}--\eqref{eq:noncausal_capacity} can be evaluated for any value of $b\in[0,\bar{B}]$ regardless of the actual value of~$B_0$.
In fact, $B_0$ can even be a random variable or an arbitrary value in $[0,\bar{B}]$, unknown to the transmitter and the receiver. By ``waiting'' a period of time before starting transmission, during which the transmitter remains silent, and which is long enough to charge the battery from 0 to $\bar{B}$, we can essentially transmit any coding scheme designed for any value of $B_0$.
\begin{proposition}
\label{prop:initial_battery_state}
The capacity of the energy harvesting channel does not depend on the initial battery state $B_0$.
\end{proposition}
See Appendix~\ref{subsec:capacity_initial_battery_state} for the proof.

We next turn our attention to the power control problem for an energy harvesting communication system that has been of significant interest in the recent communication theory literature \cite{YangUlukus2012,Ozeletal2011,TutuncuogluYener2012,online_infiniteB1,
online_infiniteB2,online_infiniteB3,online_infiniteB4,info2013,DP0,DP1,DP2,DP3, Mitran}. We introduce some new terms and notations to define the problem. A power control \emph{policy} for an energy harvesting system is a sequence of mappings from energy arrivals to a non-negative number, which will denote a level of instantaneous power.
More precisely, an \emph{online policy} $g^n$ is a sequence of mappings 
\begin{equation}\label{eq:online}
g_t:\mathcal{E}^t\to\mathbb{R}_+\qquad
,t=1,\ldots,n,
\end{equation}
and an \emph{offline policy} $g^n$ is a sequence of mappings
\begin{equation}
g_t:\mathcal{E}^n\to\mathbb{R}_+\qquad
,t=1,\ldots,n.
\end{equation}
An \emph{admissible policy} is such that satisfies the energy constraints.
%~\eqref{eq:EH_constraint} and~\eqref{eq:EH_battery}.
Formally, the set of all admissible policies with initial battery level $b$ is:
\begin{align}
\mathcal{G}_n(b)=\big\{&g^n|\ \text{s.t. }\forall e^n\in\mathcal{E}^n:\nonumber\\*
&g_t\leq b_t,\ b_0=b,\nonumber\\*
&b_t=\min\{b_{t-1}-g_{t-1}+e_t,\bar{B}\}\big\}.
	\label{eq:G_definition}
\end{align}
We denote by $\Gonline(b)$ the set of all admissible online policies, and by $\Goffline(b)$ the set of all admissible offline policies.

For a given online policy of length $n$, we define the average throughput to be:
\begin{equation}
\T(g^n)=\frac{1}{n}\mathbb{E}\left[
\sum_{t=1}^{n}\frac{1}{2}\log(1+g_t(E^t))\right],
\end{equation}
where the expectation is over the energy arrivals $E_1,\ldots,E_n$, and similarly for an offline policy, the average throughput is:
\begin{equation}
\T(g^n)=\frac{1}{n}\mathbb{E}\left[
\sum_{t=1}^{n}\frac{1}{2}\log(1+g_t(E^n))\right].
\end{equation}
Next, we define the following optimization problems which aim to maximize the long-term average throughput
\begin{align}
\Conline
&=\liminf_{n\to\infty}\max_{g^n\in\Gonline(b)}\T(g^n),
	\label{eq:online_opt}\\*
\Coffline
&=\liminf_{n\to\infty}\max_{g^n\in\Goffline(b)}\T(g^n).
	\label{eq:offline_opt}
\end{align}
Equations~\eqref{eq:online_opt} and~\eqref{eq:offline_opt} describe the online and offline power control optimization problems, respectively, studied extensively in the literature.\footnote{In the literature \cite{YangUlukus2012,Ozeletal2011,TutuncuogluYener2012}, the offline power control problem is typically studied for an arbitrary known sequence of energy arrivals, without imposing a distribution on the energy arrivals. Note that even with this difference, the resultant optimization problems and the corresponding optimal offline policies are equivalent, since the offline policy in our current case needs to maximize the throughput achieved under any given realization of the process. Imposing a distribution on the energy arrivals allows us to have a notion of long-term average offline throughput.}
In both problems, we want to maximize the long-term average throughput subject to energy constraints as given in~\eqref{eq:G_definition}, assuming there exists a transmission scheme for which allocating power $p_t$ at time $t$ yields an information rate  $r(p_t)=\frac{1}{2}\log(1+p_t)$. While the optimal offline power control policy has been explicitly characterized in \cite{YangUlukus2012,Ozeletal2011,TutuncuogluYener2012}, there is limited understanding regarding the structure of the optimal online power control policy and the resultant long-term average throughput. In the following theorem, we characterize the long-term average throughput in the online case within a constant gap independent of system parameters.

\begin{theorem}\label{thm:onlinePC}
The solutions to the online and offline power control problems are bounded by
%\begin{equation*}
%\frac{1}{2}\log(1+\mu) \geq \bar{C}^\text{offline} \geq \bar{C}^\text{online} \geq \frac{1}{2}\log(1+\mu) - 1.80
%\end{equation*}
\[
\frac{1}{2}\log(1+\mu)-1.80\leq\Conline
\leq\Coffline\leq\frac{1}{2}\log(1+\mu),
\]
where $\mu\triangleq \mathbb{E}[\min\{E_t,\bar{B}\}]$.
\end{theorem}

While the proof of the upper bound follows from a simple application of Jensen's inequality, to establish the lower bound on the throughput we construct an explicit online power control policy ${g}^n$ and show that the long-term average throughput it achieves can be at most $1.8$ bits/channel use away from the upper bound, i.e.
\begin{equation}\label{eq:gstar}
\liminf_{n\to\infty}\T({g}^n)\geq \frac{1}{2}\log(1+\mu)-1.80.
\end{equation}
This power control policy has a surprising structure: it waits for the battery to be recharged completely and then allocates power in an exponentially decaying manner. The proof of the theorem and the corresponding approximately optimal online power control policy are given in Section~\ref{sec:power_control}.\footnote{\label{foot:fixed_fraction}
In a recent publication~\cite{shaviv2015universally}, the \emph{Fixed Fraction} policy is suggested, in which $g_t=pB_t$, i.e. a fixed fraction of the battery level is allocated at each time slot. This policy is shown to achieve the optimal throughput up to a gap of only 0.72, by showing that Bernoulli energy arrivals yield the worst performance for this policy. This result combined with Theorem~\ref{thm:bounds}, which is stated next, can be immediately used to decrease the approximation gap for  $\CcausalTxRx$ and $\CnoncausalTxRx $ in Theorem~\ref{thm:capacity} to $1.77$ from $2.85$. However, it does not lead to a similar improvement for the approximation gap for $\CcausalTx$ since, as given by Theorem~\ref{thm:bounds}, the gap for $\CcausalTx$ also depends on the entropy per symbol of the online policy. Since the entropy per symbol of the policy in \cite{shaviv2015universally} is equal to $H(E_t)$, which can be arbitrarily large, here we devise an alternative online policy for which we can simultaneously have the guarantees provided in Propositions~\ref{thm:onlinePC} and \ref{prop:prop1}.
}

The next step is to connect the two problems discussed so far. In particular, we show that the solution of the power control optimization problem can be used to lower and upper bound the information-theoretic capacity of the channel in Theorem~\ref{thm:capacity}, which involves a much harder optimization problem. 

\begin{theorem}
	\label{thm:bounds}
	The capacities of the energy harvesting channel with various levels of energy arrival information can be bounded by
	\begin{subequations}
	\label{eq:causal_bounds}
	\begin{align}
	\CcausalTx&\geq
		\liminf_{n\to\infty}\max_{g^n\in\Gonline(b)}
		\{\T(g^n)-\tfrac{1}{n}H(g^n(E^n))\}\nonumber\\*
		&\hspace{12em}
		-\frac{1}{2}\log\left(\frac{\pi e}{2}\right),
		\label{eq:causal_lower_bound}\\*
	\CcausalTx &\leq\Conline,
	\label{eq:causal_upper_bound}
	\end{align}
	\end{subequations}
	\begin{IEEEeqnarray}{rLL}
		\Conline
			-\frac{1}{2}\log\left(\frac{\pi e}{2}\right)
		&\leq \CcausalTxRx &\leq\Conline,
			\label{eq:causalRx_bounds}\\
		\Coffline
			-\frac{1}{2}\log\left(\frac{\pi e}{2}\right)
		&\leq \CnoncausalTxRx &\leq\Coffline.
		\quad\ \ \label{eq:noncausal_bounds}
	\end{IEEEeqnarray} 
\end{theorem}
Note that $\frac{1}{2}\log\left(\frac{\pi e}{2}\right)\approx 1.05$. The proof is given in Section~\ref{sec:capacity_bounds}.\footnote{\label{foot:continuous_E}
Theorems~\ref{thm:main_theorem}--\ref{thm:bounds} are stated for $E_t$ that follows a discrete distribution. However, as mentioned in Section~\ref{sec:system_model}, these results can be extended to arbitrary probability distributions.
The exact derivation exceeds the scope of this paper, however, we mention briefly that this can be done by considering quantized versions of $E_t$
and applying Theorems~\ref{thm:capacity} and~\ref{thm:bounds}.
In the limit of very high resolution, it can be shown that the throughputs $\Conline$ and $\Coffline$ converge to the throughputs of the original process $E_t$.
This, along with the fact that Theorem~\ref{thm:onlinePC} holds for any distribution of $E_t$, yields Theorem~\ref{thm:main_theorem}.
}
Note that $g^n(E^n)$ in \eqref{eq:causal_lower_bound} being a deterministic (and causal) function of $E^n$ can be regarded as a random process itself and the term  $\tfrac{1}{n}H(g^n(E^n))$ corresponds to the entropy per symbol of the first $n$ symbols of this process. Note that we can further lower bound \eqref{eq:causal_lower_bound} to obtain 
\[
\Conline- H(E_t)-\frac{1}{2}\log\left(\frac{\pi e}{2}\right)\leq\CcausalTx \leq\Conline,
\]
since $g^n$ is a deterministic function of $E^n$ and therefore $H(g^n(E^n))\leq H(E^n)$. However, the original form of the lower bound in \eqref{eq:causal_lower_bound} can be significantly tighter than the form above since the entropy of the allocated power process~$g^n(E^n)$ can be significantly smaller than the entropy of the energy harvesting process $E^n$. In particular, we show in Section~\ref{sec:power_control} that the online power control policy we develop to achieve the lower bound in Theorem~\ref{thm:onlinePC} has entropy per symbol bounded by $1$ bit/channel use. This is formally stated in the following proposition. The form of the lower bound in \eqref{eq:causal_lower_bound} (and more specifically its proof) reveal a trade-off in designing communication strategies for channels with state information available only at the transmitter: while the transmitter knowing the state of the channel can follow a different strategy for each value of the state, the need to infer the state from the received signal can lead to a rate hit proportional to the entropy of the state. This can make strategies that have a coarser dependence on the state more desirable.

\begin{proposition}\label{prop:prop1}
For the online power control policy ${g}^n$ achieving \eqref{eq:gstar}, 
we have 
\[
\frac{1}{n}H({g}^n(E^n))\leq 1.
\] 
In particular, this property of ${g}^n$ together with \eqref{eq:gstar} yields
\begin{align}
\lefteqn{
\liminf_{n\to\infty} 
\max_{g^n\in\Gonline(b)}
\left\{\T(g^n)-\tfrac{1}{n}H(g^n(E^n))\right\}}
\nonumber\\*
&\hspace{10em} \geq \frac{1}{2}\log(1+\mu) - 2.80.
\label{eq:lowerbound}
\end{align}
\end{proposition}
It is immediate to verify this proposition, which we do in Section~\ref{sec:power_control}, after we introduce the policy $g^n$.

Note that when the receiver has knowledge of the energy arrival process in \eqref{eq:causalRx_bounds} and \eqref{eq:noncausal_bounds}, the gap between the information-theoretic capacity and the long-term average throughput is only $\frac{1}{2}\log\left(\frac{\pi e}{2}\right)$  which is approximately 1.05.
It is indeed surprising that the actual information-theoretic capacity achieves, within a constant gap, the solution of the power control problem. In Section~\ref{sec:capacity_bounds} we suggest one natural way to use the optimal power control policy to construct explicit codes which achieve the lower bounds in \eqref{eq:causal_bounds}--\eqref{eq:noncausal_bounds}. This emphasizes the importance of the power control problem %\cite{YangUlukus2012,Ozeletal2011,TutuncuogluYener2012} 
in  understanding the more fundamental information-theoretic problem.

%Equations~\eqref{eq:online_opt} and~\eqref{eq:offline_opt} describe the online and offline power control optimization problems, respectively, studied extensively in the literature (e.g.~\cite{YangUlukus2012,Ozeletal2011,TutuncuogluYener2012}).
%In both problems, we want to maximize the long-term average information rate subject to energy constraints as given in~\eqref{eq:G_definition} and~\eqref{eq:H_definition}, assuming there exists a transmission scheme for which allocating power $p_t$ at time $t$ yields an information rate  $r(p_t)=\frac{1}{2}\log(1+p_t)$. 

The additive approximations in Theorem~\ref{thm:bounds} obviously become irrelevant at low-SNR since the lower bounds~\eqref{eq:causal_bounds}-\eqref{eq:noncausal_bounds} can become negative, and thus useless. This indeed is an artifact of our constant gap approximation approach which bounds the worst case additive gap between the quantities of interest (the worst case gap for all the approximation theorems above occurs typically when $\mu$ and $\bar{B}$ are very large). It should be clear from the proofs of these approximation theorems that as the quantities of interest become small, the additive gap between them also tends to zero. In order to illustrate this fact, in the following theorem we 
provide a multiplicative relation between the information-theoretic capacity and the long-term average throughput which is more relevant in the low-SNR regime.
\begin{theorem}%[Multiplicative Bounds]
	\label{thm:mult_bounds}
	The capacities of the energy harvesting channel with various levels of energy arrival information can be bounded by
	\begin{subequations}
	\label{eq:mult_bound_causal}
	\begin{align}
	\CcausalTx&\geq\liminf_{n\to\infty}
		\max_{g^n\in\Gonline(b)}
		\{\eta\cdot\T(g^n)
		-\tfrac{1}{n}H(g^n(E^n))\},
		\label{eq:mult_lower_bound_causal}\\
	\CcausalTx&\leq\Conline,
		\label{eq:mult_upper_bound_causal}
	\end{align}
	\end{subequations}
	\begin{IEEEeqnarray}{rLL}
%		\lim_{n\to\infty}
%		\max_{g^n\in\mathcal{G}_n^{\mathrm{online}}}
%		\{\eta\cdot\T(g^n)
%		-\tfrac{1}{n}H(g^n(E^n))\}
%		&\leq \CcausalTx &\leq\Conline,
%			\label{eq:mult_bound_causal}\\
		\eta\cdot\Conline
		&\leq \CcausalTxRx &\leq\Conline,
			\label{eq:mult_bound_causalRx}\\
		\eta\cdot\Coffline
		&\leq \CnoncausalTxRx &\leq\Coffline,
		\quad\ \ \label{eq:mult_bound_noncausal}
	\end{IEEEeqnarray} 
where the parameter $\eta\geq 0.7473$.
\end{theorem}
See Appendix~\ref{sec:mult_bounds} for the proof.
Again, note that~\eqref{eq:mult_lower_bound_causal} can be further lower bounded to obtain
\[
\eta\cdot\Conline-H(E_t)\leq\CcausalTx\leq\Conline.
\]

Finally, note that combining \eqref{eq:lowerbound} with the inequalities in Theorem~\ref{thm:bounds}, we immediately obtain Theorem~\ref{thm:main_theorem}.

%---------------------------%
\section{Channel Capacity: Proof of Theorem~\ref{thm:capacity}}
%---------------------------%
\label{sec:capacity_proof}

We begin with proof of achievability for the first case~-- causal energy arrival information at the transmitter alone, and we consider the equivalent channel model developed in the beginning of Section~\ref{sec:equivalent_channel_and_main_results}.
We construct an achievable scheme composed of $k$ blocks with each block containing a codeword of length $n$ which is an element of $\mathcal{U}^n$.
As such, each codeword is a function of only the past $n$ energy arrivals, which means we ignore information regarding all the energy arrivals in the previous blocks.
These codewords are designed to satisfy the energy constraints for initial battery level $B_0=b$, so to accommodate this, we must ensure that the battery level in the beginning of each block is at least $b$.
To this matter, we allow the battery to ``recharge'' after we transmit each codeword by waiting a sufficient amount of time ($\ell$ time slots), during which the transmitter remains silent.
If $\ell$ is large enough, the probability of recharging the battery back to level $b$ will be high.
In the case when the battery is not sufficiently recharged at the beginning of the next block, we can simply give up on this block and transmit the all-zero codeword. We will explicitly show that this will have a negligible effect on the achievable rate.

To make the probabilistic analysis simpler, it is helpful to have the different blocks statistically independent of each other.
Note that subsequent blocks are coupled through the battery state. More precisely,  the event of whether the battery at the beginning of each block is recharged to $b$ or not (which, in turn, determines whether a codeword is being sent or just zeros) may depend on the amount of energy left in the battery at the end of the transmission in the previous block.
To decouple one block from its sequel, we \emph{purposely deplete} the battery to zero before waiting for it to be recharged.
This way, the battery level at the beginning of the next block will depend solely on the last $\ell$ energy arrivals.
In what follows, we make the above description precise.
\footnote{Strictly speaking, there are no energy arrivals and battery state in the equivalent model defined in \eqref{eq:inputalphabet}-\eqref{eq:U_battery}, but $e^t$ and $b_t$ in \eqref{eq:U_battery} can be rather viewed as dummy variables where $e^t$ represents the input variables of the function $u^t$ and $b_t$ is an intermediate variable used to define the input constraint. However, we continue to refer to $e^t$ as the energy arrival sequence up to time $t$ and $b_t$ as the battery state at time $t$ to make the exposition easier.}
\footnote{The idea of ``erasing'' the memory in the battery by using codewords interleaved with silent times has first appeared in~\cite{JogAnantharam2014} which considers a special case of the problem with constant deterministic energy arrivals. A block i.i.d. coding scheme was proposed in \cite{MaoHassibi2014} when the transmitter has causal energy arrival information but  $b=0$, in which case one does not need the zero-padding between the codewords to recharge the battery. Our achievable strategy is closer to \cite{Ozeletal2014} which considers noncausal energy arrival information at the transmitter in the case of $B_0=\bar{B}$. However, the proof in \cite{Ozeletal2014} is incomplete because it assumes that by making the zero padding between blocks long enough, we can ensure that the battery is recharged to full each and every time. This is not possible, because as the number of blocks $k\to\infty$, recharging failures are inevitable and have to be explicitly taken into account.  Also, in the noncausal case, the codewords can be constructed directly on the original channel with input alphabet $\mathcal{X}$ and not $\mathcal{U}^n$ as we do here (See Appendix~\ref{sec:capacity_noncausal_Rx}).
In \cite{ShavivOzgur2015}, we show that the noncausal capacity is strictly larger and therefore different than the causal capacity.}

%Fix integers $n$ and $\ell$.
Fix $P_{U^n}\in\mathcal{P}_n(b)$ and for each message $w$, generate $k$ random codewords independently $\mathbf{v}_i\sim P_{U^n}$, $i=1,\ldots,k$. Recall that each $\mathbf{v}_i$ is a function on $\mathcal{E}^n$. The chosen message $w$ will be transmitted over $k$ blocks, each of size $n+\ell+1$, for a total transmit time of $k(n+\ell+1)$.
Hence, we will define codewords $u^{k(n+\ell+1)}\in\mathcal{U}^{k(n+\ell+1)}$ using the above $\mathbf{v}_i$.

Each block comprises three parts: the first part, of length $n$, consists of the codeword $\mathbf{v}_i$ (or an all-zero vector of length $n$ if the battery level at the beginning of this block is not sufficient to transmit codeword $\mathbf{v}_i$). In the second part, which takes only one time slot, we deplete the battery to $0$. The third part consists of $\ell$ zeros, which are meant to recharge the battery to level $b$.

Consider block $i$, $1\leq i\leq k$, which takes place during times $t=(i-1)(n+\ell+1)+1$ to $t=i(n+\ell+1)$.
We define the following notations:
$b_{0,i}\triangleq b_{(i-1)(n+\ell+1)}$
is the battery level before the beginning of the block, i.e. the initial battery level before we start transmitting codeword $\mathbf{v}_i$;
$z_i\triangleq\mathbf{1}_{\{b_{0,i}\geq b\}}$
is an indicator, denoting whether the initial battery level $b_{0,i}$ is sufficient to transmit the codeword $\mathbf{v}_i$;
$b_{d,i}\triangleq b_{(i-1)(n+\ell+1)+n+1}$
is the battery level at time $(i-1)(n+\ell+1)+n+1$,  which will be used to deplete the battery at this time-slot;
$\mathbf{e}_i\triangleq e_{(i-1)(n+\ell+1)+1}^{(i-1)(n+\ell+1)+n}$
are the energy arrivals during the transmission of the codeword $\mathbf{v}_i$;
and $\tilde{\mathbf{e}}_i\triangleq e_{(i-1)(n+\ell+1)+n+2}^{i(n+\ell+1)}$
are the energy arrivals during the zero-padding phase at the end of each block.
Denote by $\mathbf{0}$ a vector of $\ell$ zeros.

%The scheme operates as follows:
Block $i$ of the codeword is constructed as follows:
at the beginning of block $i$ at time $t=(i-1)(n+\ell+1)+1$, given $u^{t-2}$ for each $e^{t-1}$ we can compute $b_{0,i}=b_{t-1}$.
If the battery state is at least $b$, we send the codeword $\mathbf{v}_i$, i.e. $u_t^{t+n-1}(e^{t+n-1})=\mathbf{v}_i(\mathbf{e}_i)$.
Otherwise, the transmitter sends zeros for $n$ time slots.
Thus, the first part of block $i$ can be written as $z_i\cdot\mathbf{v}_i$.
For the second part of the block, which is the single time-slot $t+n$, again  given $u^{t+n-1}$ for each $e^{t+n}$ the transmitter computes the battery state $b_{d,i}=b_{t+n}$, and transmits $\sqrt{b_{d,i}}$, i.e. $u_{t+n}(e^{t+n})=\sqrt{b_{d,i}(e^{t+n})}$.
This will deplete the battery to zero.
By purposely depleting the battery before recharging it, we remove the dependence between different blocks, and make the probability of recharge failure an i.i.d. process.
Next, the transmitter sends zeros for $\ell$ time slots, which will recharge the battery to $b$ with high probability.

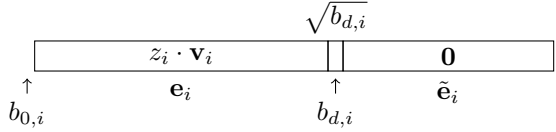
\begin{figure}[!t]
\centering
\begin{tikzpicture}
	\draw (0.1,0) rectangle (4,0.4) 
		node[pos=.5] {$z_i\cdot\mathbf{v}_i$}
		node[below=7,pos=.5] {$\mathbf{e}_i$};
	\draw (4,0) rectangle (4.2,0.4)
		node[above=5,pos=.5] {$\sqrt{b_{d,i}}$};
	\draw (4.2,0) rectangle (7,0.4) 
		node[pos=.5] {$\mathbf{0}$} 
		node[below=6,pos=.5] {$\tilde{\mathbf{e}}_i$};
	
	\tikzstyle{every path}=[draw,->]
	\draw[->] (0,-0.3) -- node[below,pos=0.1] {$b_{0,i}$} (0,-0.1);
	\draw[->] (4.1,-0.3) -- node[below,pos=0.1] {$b_{d,i}$} (4.1,-0.1);
\end{tikzpicture}
\caption{Structure of block $i$ in the coding scheme.}
\label{fig:block_structure}
\end{figure}

To summarize, the transmitted block is
\begin{equation}
\label{eq:u_block_def}
	u_{(i-1)(n+\ell+1)+1}^{i(n+\ell+1)}
		\big(e^{i(n+\ell+1)}\big)
	=\big[z_i\cdot\mathbf{v}_i(\mathbf{e}_i),\ 
	\sqrt{b_{d,i}},\ \mathbf{0}\big],
\end{equation}
See Fig.~\ref{fig:block_structure} for a graphical representation of the block structure.
Note that since the battery is depleted to zero at the second part of block $i$, $z_{i+1}$ is a deterministic function of $\tilde{\mathbf{e}}_i$.
In fact, $z_{i+1}$ is a deterministic function of $b_{0,i+1}$, where $b_{0,i+1}=\min\{\bar{B},\sum_{t=(i-1)(n+\ell+1)+n+2}^{i(n+\ell+1)}e_t\}$.

$u^{k(n+\ell+1)}(e^{k(n+\ell+1)})$ defined in~\eqref{eq:u_block_def} is a well-defined element in $\mathcal{U}^{k(n+\ell+1)}$.
Moreover, observe that it satisfies the energy constraint:
this is trivial for $u^{n+\ell+1}$.
For the subsequent codewords, 
if the battery level is larger than $b$, we assume it is $b$ and ignore (waste) the remaining energy.
If it is less than $b$, we transmit only zeros.
This will satisfy the energy constraints.
%For a rigorous proof that the constraints are satisfied, see Appendix~\ref{subsec:concatenated_codewords}.

%Denote the physical channel input at the first part of block $i$ by $\mathbf{x}_i=x_{(i-1)(n+\ell+1)+1}^{(i-1)(n+\ell+1)+n}$, and the channel output $\mathbf{y}_i=y_{(i-1)(n+\ell+1)+1}^{(i-1)(n+\ell+1)+n}$.
Denote the channel output during the first part of block $i$ by $\mathbf{y}_i=y_{(i-1)(n+\ell+1)+1}^{(i-1)(n+\ell+1)+n}$.
The receiver observes $y^{k(n+\ell+1)}$ but makes use only of $\mathbf{y}^k=(\mathbf{y}_1,\ldots,\mathbf{y}_k)$ for decoding, by applying standard jointly typical decoding with $\mathbf{v}^k$.
The channel transition probability from $\mathbf{V}^k$ to $\mathbf{Y}^k$ is
\begin{align*}
	\hspace{1em}\lefteqn{\hspace{-1em}
	P_{\mathbf{Y}^k|\mathbf{V}^k}(\mathbf{y}^k|\mathbf{v}^k)
	}\\*
	&=\sum_{e^{k(n+\ell+1)}}P_{E^{k(n+\ell+1)}}(e^{k(n+\ell+1)})\\*
	&\qquad\qquad\qquad\cdot
		P_{Y^{kn}|X^{kn}}\big(\mathbf{y}^k|
			z^k(\tilde{\mathbf{e}}^{k-1})
			\cdot \mathbf{v}^k(\mathbf{e}^k)\big)\\
	&=\sum_{e^{k(n+\ell+1)}}
		\prod_{i=1}^{k}P_{E^n}(\mathbf{e}_i)P_E(e_{(i-1)(n+\ell+1)+n+1})
			P_{E^\ell}(\tilde{\mathbf{e}}_i)\\*
		&\qquad\qquad\qquad\cdot
			P_{Y^n|X^n}\big(\mathbf{y}_i|
			z_i(\tilde{\mathbf{e}}_{i-1})\cdot
			\mathbf{v}_i(\mathbf{e}_i)\big)\\
	&=\prod_{i=1}^{k}\sum_{\mathbf{e}_i,\tilde{\mathbf{e}}_i}
		P_{E^n}(\mathbf{e}_i)P_{E^\ell}(\tilde{\mathbf{e}}_{i})
		P_{Y^n|X^n}\big(\mathbf{y}_i|
		z_i(\tilde{\mathbf{e}}_{i-1})\cdot
		\mathbf{v}_i(\mathbf{e}_i)
		\big)\\
	&=\prod_{i=1}^{k}\sum_{\mathbf{e}_i,z_i}
		P_{E^n}(\mathbf{e}_i)P_Z(z_i)
		P_{Y^n|X^n}(\mathbf{y}_i|
		z_i\cdot\mathbf{v}_i(\mathbf{e}_i)).
\end{align*}
Note that since $E_t$ is i.i.d., $P_{E^n}$ and $P_Z$ do not depend on $i$,\footnote{
$Z_1$ is an exception, since it equals 1 w.p. 1. Nevertheless, one can artificially generate a Bernoulli RV and choose whether to transmit $\mathbf{v}_1$ or zeros according to the outcome.
}
so this is a memoryless channel with transition probability
\begin{align*}
	P_{\mathbf{Y}|\mathbf{V}}(\mathbf{y}|\mathbf{v})
	&=\sum_{e^n,z}P_{E^n}(e^n)P_Z(z)
	P_{Y^n|X^n}\big(\mathbf{y}|z\cdot
	\mathbf{v}(e^n)\big)\\
	&=\sum_zP_Z(z)P_{Y^n|U^n}(\mathbf{y}|z\cdot\mathbf{v}).
\end{align*}
where the last step is from~\eqref{eq:U_transition_prob}.
Note that $\mathbf{Y}=Y^n$ is the output of the channel from $U^n$ to $Y^n$ with the input multiplied by an independent Bernoulli RV $Z$.

Denote $P_Z(0)=\alpha$.
Taking $k\to\infty$, we get by standard joint typicality arguments that rate $I(\mathbf{V};\mathbf{Y})$ is achievable.
The following holds:
\begin{align*}
	I(\mathbf{V};\mathbf{Y})
	&=I(\mathbf{V};\mathbf{Y},Z)-I(\mathbf{V};Z|\mathbf{Y})\\
	&\geq I(\mathbf{V};\mathbf{Y}|Z)-H(Z)\\
	&=(1-\alpha)I(\mathbf{V};\mathbf{Y}|Z=1)-H_2(\alpha),
\end{align*}
where $H_2(\cdot)$ is the binary entropy function.
The last step is because $I(\mathbf{V};\mathbf{Y}|Z=0)=0$.

Note that $P_{\mathbf{Y}|\mathbf{V},Z}(\mathbf{y}|\mathbf{v},{z=1})=P_{Y^n|U^n}(\mathbf{y}|\mathbf{v})$
and, by construction, $\mathbf{V}\sim P_{U^n}$ independent of $Z$.
This implies
\[
	I(\mathbf{V};\mathbf{Y}|Z=1)=I(U^n;Y^n).
\]
For $\ell$ large enough, $\alpha$ can be upper bounded using the law of large numbers
$
	\alpha=\Pr\{\sum_{t=1}^{\ell}E_t<b\}
	\leq\varepsilon_\ell,
$
where $\lim_{\ell\to\infty}\varepsilon_\ell=0$
(recall that $\mathbb{E}[E_t]>0$),
s.t. for every $n\geq1$ we have
\[
	\CcausalTx\geq
	\frac{(1-\varepsilon_\ell)I(U^n;Y^n)
		-H_2(\varepsilon_\ell)}
		{n+\ell+1}.
\]
Since $P_{U^n}$ is an arbitrary input distribution in $\mathcal{P}_n(b)$, we can take the supremum to obtain
\[
	\CcausalTx\geq\sup_{P_{U^n}\in\mathcal{P}_n(b)}
	\frac{(1-\varepsilon_\ell)I(U^n;Y^n)
		-H_2(\varepsilon_\ell)}
		{n+\ell+1}.
\]
Let $\ell=\lceil\log n\rceil$.
Taking $n\to\infty$, we get
\begin{align}
	\CcausalTx
	&\geq\limsup_{n\to\infty}\frac{1}{n}
		\sup_{P_{U^n}\in\mathcal{P}_n(b)}I(U^n;Y^n).
		\label{eq:EH_achievability}
\end{align}

For the converse part, we use Fano's inequality as in~\cite{VerduHan1994}.
%For completeness, this is shown in Appendix~\ref{sec:converse}.
%We bring the converse to Theorem~\ref{thm:capacity}, as shown in~\cite{VerduHan1994}.
An $(M,n,\varepsilon)$ code for the channel defined in Section~\ref{sec:equivalent_channel_and_main_results} satisfies
\begin{align*}
	H(W|Y^n)&\leq H_2(\varepsilon)+\varepsilon\log M\\
	(1-\varepsilon)\log M&\leq I(W;Y^n)+H_2(\varepsilon)
\end{align*}
If $R<\CcausalTx$ is achievable, then for every $\delta>0$,
\[
	R-\delta<\frac{1}{n}\frac{1}{1-\varepsilon}
		[I(U^n;Y^n)+H_2(\varepsilon)],
\]
where $I(U^n;Y^n)$ is the mutual information evaluated for $P_{U^n}$ induced by the code.
Since all codewords must satisfy the input constraints~\eqref{eq:U_energy} and~\eqref{eq:U_battery}, this implies $P_{U^n}\in\mathcal{P}_n(b)$ (see~\eqref{eq:U_constraint}).
Therefore
\[
	R-\delta<\frac{1}{n}\frac{1}{1-\varepsilon}
	\left[\sup_{P_{U^n}\in\mathcal{P}_n(b)}I(U^n;Y^n)
	+H_2(\varepsilon)\right],
\]
which implies
\[
	R\leq\frac{1}{1-\varepsilon}
	\liminf_{n\to\infty}\frac{1}{n}
	\sup_{P_{U^n}\in\mathcal{P}_n(b)}I(U^n;Y^n).
\]
Taking $\varepsilon\to0$ gives 
\begin{equation}
	\CcausalTx\leq\liminf_{n\to\infty}\frac{1}{n}
	\sup_{P_{U^n}\in\mathcal{P}_n(b)}I(U^n;Y^n).
	\label{eq:EH_converse}
\end{equation}
Together with~\eqref{eq:EH_achievability}, this implies that the limit exists and is given by~\eqref{eq:causal_capacity}.

We now turn to the case of energy arrival information available causally at the transmitter and the receiver (eq.~\eqref{eq:causalRx_capacity}). We can repeat the previous steps in exactly the same manner, and since the receiver now observes $E^n$ as well, we simply add it alongside $Y^n$.
All the arguments still hold, and we are left with 
\begin{align*}
	\CcausalTxRx&=\lim_{n\to\infty}\frac{1}{n}
		\sup_{P_{U^n}\in\mathcal{P}_n(b)}
		I(U^n;Y^n,E^n)\\
	&\overset{\text{(i)}}{=}
		\lim_{n\to\infty}\frac{1}{n}
		\sup_{P_{U^n}\in\mathcal{P}_n(b)}
		I(U^n;Y^n|E^n)\\
	&\overset{\text{(ii)}}{=}
		\lim_{n\to\infty}\frac{1}{n}
		\sup_{P_{U^n}\in\mathcal{P}_n(b)}
		I(X^n;Y^n|E^n)\\
	&\overset{\text{(iii)}}{=}
		\lim_{n\to\infty}\frac{1}{n}
		\sup_{P_{X^n\|E^n}\in\mathcal{Q}_n(b)}
		I(X^n;Y^n|E^n).
\end{align*}
where (i) is because $U^n$ is independent of $E^n$; (ii) is because $X^n=U^n(E^n)$ and the Markov chain $U^n-(X^n,E^n)-Y^n$;
and (iii) is because, as will be shown below in~\eqref{eq:strategies_induce_causally_conditioned}, any distribution $P_{U^n}\in\mathcal{P}_n(b)$ induces a distribution $P_{X^n\|E^n}\in\mathcal{Q}_n(b)$ on $X^n$, and any $X^n\sim P_{X^n\|E^n}\in\mathcal{Q}_n(b)$ can be represented as a random function of $E^n$ according to some distribution $P_{U^n}\in\mathcal{P}_n(b)$.
This gives~\eqref{eq:causalRx_capacity}.

To show (iii), observe that
the joint distribution $P_{X^n,E^n,U^n}$ can be factored as
\begin{align*}
&P_{X^n,E^n,U^n}(x^n,e^n,u^n)\nonumber\\*
&\hspace{2em}
=\prod_{i=1}^{n}
P_{U_i|U^{i-1}}(u_i|u^{i-1})P_{E}(e_i)1\{x_i=u_i(e^i)\}.
\end{align*}
Summing over $x_n$ and then over $u_n$ gives
\begin{align*}
&P_{X^{n-1},E^n,U^{n-1}}(x^{n-1},e^n,u^{n-1})\nonumber\\*
&\ =\prod_{i=1}^{n-1}P_{U_i|U^{i-1}}(u_i|u^{i-1})P_E(e_i)1\{x_i=u_i(e^i)\}\cdot P_E(e_n).
\end{align*}
Continuing to sum over $x_{n-1},u_{n-1}$, and then $x_{n-2},u_{n-2}$ and so forth yields for any $t$:
\begin{align*}
&P_{X^t,E^n,U^t}(x^t,e^n,u^t)\nonumber\\*
&\ =\prod_{i=1}^{t}P_{U_i|U^{i-1}}(u_i|u^{i-1})1\{x_i=u_i(e^i)\}\cdot \prod_{i=1}^{n}P_E(e_i).
\end{align*}
Summing over $e_{t+1}^n$ gives
\begin{align*}
&P_{X^t,E^t,U^t}(x^t,e^t,u^t)\nonumber\\*
&\quad =\prod_{i=1}^{t}P_{U_i|U^{i-1}}(u_i|u^{i-1})P_E(e_i)1\{x_i=u_i(e^i)\}
,
\end{align*}
hence
\[
P_{X^t,E^n,U^t}(x^t,e^n,u^t)=P_{X^t,E^t,U^t}(x^t,e^t,u^t)\cdot\prod_{i=t+1}^{n}P_E(e_i).
\]
Summing over $u^t$ gives 
\[
P_{X^t,E^n}(x^t,e^n)=P_{X^t,E^t}(x^t,e^t)\prod_{i=t+1}^{n}P_E(e_i).
\]
This implies 
\[
P_{X^{t-1},E^t}(x^{t-1},e^t)=P_{X^{t-1},E^{t-1}}(x^{t-1},e^{t-1})P_E(e_t).
\]
Using all of the above identities, we have:
\begin{align}
\hspace{3em}\lefteqn{\hspace{-3em}P_{X_t|X^{t-1},E^n}(x_t|x^{t-1},e^n)}\nonumber\\*
&=\frac{P_{X^t,E^n}(x^t,e^n)}{P_{X^{t-1},E^n}(x^{t-1},e^n)}\nonumber\\
&=\frac{P_{X^t,E^t}(x^t,e^t)\prod_{i=t+1}^{n}P_E(e_i)}{P_{X^{t-1},E^{t-1}}(x^{t-1},e^{t-1})\prod_{i=t}^{n}P_E(e_i)}\nonumber\\
&=\frac{P_{X^t,E^t}(x^t,e^t)\prod_{i=t+1}^{n}P_E(e_i)}{P_{X^{t-1},E^t}(x^{t-1},e^{t})\prod_{i=t+1}^{n}P_E(e_i)}\nonumber\\
&=P_{X_t|X^{t-1},E^t}(x_t|x^{t-1},e^t).
\label{eq:strategies_induce_causally_conditioned}
\end{align}
This implies $P_{X^n|E^n}(x^n|e^n)=P_{X^n\|E^n}(x^n\|e^n)$.

The proof of the remaining case (namely energy arrival information available at the transmitter and the receiver noncausally) follows exactly the same lines and appears in Appendix~\ref{sec:capacity_noncausal_Rx}.

\section{Optimal Online Power Control: Proof of Theorem \ref{thm:onlinePC}}
\label{sec:power_control}

In this section, we consider the power control problem for the energy harvesting communication system. This problem was formally defined in \eqref{eq:online}-\eqref{eq:offline_opt} in the two settings of interest, offline and online. Here we focus on the online version of the problem. To recall, the goal is to find an optimal online power control policy 
\begin{equation*}
g_t:\mathcal{E}^t\to\mathbb{R}_+\qquad
,t=1,\ldots,n,
\end{equation*}
that satisfies the energy constraints, i.e. it belongs to the set
\begin{align*}
\Gonline(b)=\big\{&g^n|\ \text{s.t. }\forall e^n\in\mathcal{E}^n:\nonumber\\*
&g_t\leq b_t,\ b_0=b,\nonumber\\*
&b_t=\min\{b_{t-1}-g_{t-1}+e_t,\bar{B}\}\big\}.
\end{align*}
and maximizes the long-term average throughput of the system
\begin{equation}
\Conline
=\liminf_{n\to\infty}\max_{g^n\in\Gonline(b)}\T(g^n).
\end{equation}
We start by deriving a simple upper bound on the throughput which serves as a benchmark for the online polices we construct in the rest of the section.

\subsection{Upper Bounding the Throughput}
The upper bound we develop in this section not only holds for~$\Conline$, which is of main interest in this section, but also for~$\Coffline$ defined in~\eqref{eq:offline_opt}. First note that without loss of generality, we can replace the random process $E_t$ with $\tilde{E}_t=\min\{E_t,\bar{B}\}$ without changing the system. This is due to the fact that whenever an energy arrival $E_t$ is larger than $\bar{B}$, it will be clipped to at most $\bar{B}$. 
Denote $\mu=\mathbb{E}[\tilde{E}_t]$. For any $n$ and any policy $g^n$, we have:
\begin{align*}
\T(g^n)
&=
	\frac{1}{n}\sum_{t=1}^n \mathbb{E}
	\left[\frac{1}{2}\log
	\big(1+g_t(\tilde{E}^n)\big)\right]\\
&\overset{\text{(i)}}{\leq}
	\frac{1}{2}\log\left(1+\frac{1}{n}
	\mathbb{E}\big[\sum_{t=1}^{n} 
	g_t(\tilde{E}^n)\big]\right)\\
&\overset{\text{(ii)}}{\leq}
	\frac{1}{2}\log\left(1+ \frac{1}{n}\mathbb{E}
	\big[\bar{B}+\sum_{t=1}^{n} \tilde{E}_t\big]
	\right)\\
&=\frac{1}{2}\log\left(1+ \frac{1}{n}\bar{B}+ \mu\right)
\end{align*}
where (i) is by concavity of $\log$; (ii) follows from the fact that the total allocated energy up to time $n$, can not exceed the total energy that arrives up to time $n$ plus the energy initially available in the battery,
\[
\sum_{t=1}^{n} g_t\leq \bar{B}+\sum_{t=1}^{n} \tilde{E}_t.
\]
The last term tends to $\frac{1}{2}\log(1+\mu)$ as $n\to\infty$. Note that this is true for both offline and online power control policies, and for any energy arrival process $E_t$. We therefore have:
\begin{equation}
\Conline\leq\Coffline\leq\frac{1}{2}\log(1+\mu),
\label{eq:power_allocation_upper_bound}
\end{equation}
where $\mu=\mathbb{E}[\min\{E_t,\bar{B}\}]$, which proves the upper bound in Theorem~\ref{thm:onlinePC}.

\subsection{Approximately Optimal Online Power Control Policies}
We next turn to developing explicit online power control policies that can provably approach the upper bound developed in the previous section. While we are interested in policies that perform well for any arbitrary i.i.d. energy harvesting process, our development is inspired by the approximately optimal online power control policy for i.i.d. Bernoulli energy arrivals developed in \cite{DongFarniaOzgur2015}. We next overview this policy and prove its approximate optimality in a somewhat simpler manner which also leads to a slightly better gap (0.721 bits/channel use as opposed to 0.973 bits/channel use in \cite{DongFarniaOzgur2015}). Our analysis for the general case leverages on this derivation.

\subsubsection{Bernoulli Energy Arrivals}\label{subsubsec:Bernoulli}
\begin{figure}
\centering
\begin{tikzpicture}
	\tikzstyle{every node}=[font=\scriptsize];
	
	\foreach \x/\g/\i in {0/1/1, 0.5/0.5/2, 1/0.25/3, 1.5/0.125/4, 2/1/1, 2.5/0.5/2, 3/0.25/3, 3.5/1/1, 4/0.5/2}
	{
		\FPeval{\mycolor}{round(100*\g,0)};
		\filldraw[fill=gray!\mycolor!white,draw=black] (\x,0) -- (\x+.5,0) -- (\x+.5,\g) -- (\x,\g) -- cycle;
		\node[above] at (\x+.25,\g) {$g_{\i}$};
	}
	
	\node at (-.5,.7) {$g_t$};	
	
	\def \h {2};
	
	\foreach \x/\b/\e/\i/\mygreen/\myyellow in {0/2/\bar{B}/1/100/100, 0.5/1/0/2/0/0, 1/0.5/0/3/0/100, 1.5/0.25/0/4/30/100, 2/2/\bar{B}/5/100/100, 2.5/1/0/6/0/0, 3/0.5/0/7/40/100, 3.5/2/\bar{B}/8/100/100, 4/1/0/9/0/40}
	{
		\FPeval{\mycolor}{round(50*\b,0)};
		\filldraw[fill=gray!\mycolor!white,draw=black] (\x,\h) -- (\x+.5,\h) -- (\x+.5,\h+\b) -- (\x,\h+\b) -- cycle;
	}
	
	\foreach \x in {0,2,3.5}
	{
		\draw[->] (\x+.25,\h+2.8) node[above] {$\bar{B}$} -- (\x+.25,\h+2.4);
	}
	
	\node[above] at (-.5,\h+2.8) {$E_t$};
	\node at (-.5,\h+1) {$B_t$};
\end{tikzpicture}
\caption{The approximately optimal online power control policy for Bernoulli energy arrivals.}
\label{fig:Bernoullipowercontrol}
\end{figure}
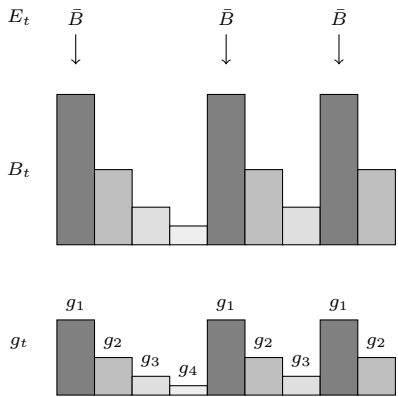

Assume the energy arrivals $E_t$ are i.i.d. Bernoulli RVs:
\[
	E_t=
	\begin{cases}
		\bar{B}&\text{w.p. }p\\
		0&\text{w.p. }1-p,
	\end{cases}
\]
i.e. at each time $t$ either the battery is fully charged to $\bar{B}$ with probability $p$ or no energy is harvested at all with  probability $1-p$. \cite{DongOzgur2014} proposes the following online power control policy for this system: Let $j_t(E^t)$ be the time of the last energy arrival, i.e.
\[
j_t(E^t)=\{\sup\ \tau\leq t:\ E_{\tau}=\bar{B}\}.
\]
The policy is defined as follows:
\[
g_t(E^t)=\bar{B}p(1-p)^{t-j_t}.
%=\mu(1-p)^{t-T}.
\]
With this policy, the amount of energy we allocate to each time slot decreases exponentially with the time since the last battery recharge (or equivalently energy arrival). Note that this is clearly an admissible strategy since 
$$
\sum_{k=j_t}^\infty\bar{B}p(1-p)^{k-j_t}=\bar{B},
$$
i.e. the total energy we allocate until the next battery recharge can never exceed $\bar{B}$, the amount of energy initially available in the battery. Another way to view this strategy is that we always use $p$ fraction of the remaining
energy in the battery at each time. Note that the energy in the battery
decays like $B_t= (1 - p)^{t-j_t}\bar{B}$. The motivation for this
power control policy can be understood as follows: for the Bernoulli
arrival process $E_t$, the inter-arrival time is a Geometric random
variable with parameter $p$. We know that the Geometric
random variable is memoryless and has mean $1/p$. Therefore,
at each time step, the expected number of time steps to the next
energy arrival is always $1/p$. Since $\log(\cdot)$ is a concave
function, we would ideally want to allocate the energy as uniformly as
possible over time, i.e. if the current energy level
in the battery is $b_t$ and we knew that the next recharge of
the battery would be in exactly $m$ channel uses, we would
allocate $b_t/m$ energy to each of the next $m$ channel uses.
For the online case of interest here, we do not know when
the next energy arrival will be. Instead, we use the expected
time to the next energy arrival: since at each time
step, the expected time to the next energy arrival is $1/p$, we
allocate a fraction $p$ of the currently available energy in the battery. 
Fig.~\ref{fig:Bernoullipowercontrol} illustrates this power control policy.

Before moving forward to establish the approximate optimality of this power control policy, we provide a few definitions and results from renewal theory.
\begin{definition}\label{def:regenerative}
A stochastic process $\{X_t\}_{t=1}^{\infty}$ is called a \emph{non-delayed regenerative process} if there exists a random time $\tau>0$ such that the process $\{X_{\tau+t}\}_{t=1}^{\infty}$ has the same distribution as $\{X_t\}_{t=1}^{\infty}$ and is independent of the past $(\tau,X^{\tau})$.
\end{definition}
Observe that a regenerative process is composed of i.i.d. ``cycles'' or \emph{epochs}, which have i.i.d. durations $\tau_1,\tau_2,\ldots$. At the beginning of each epoch, the process ``regenerates'' and all memory of the past is essentially erased.
The following theorem establishes an important time-average property of regenerative processes.
\begin{theorem}[LLN for Regenerative Processes]
\label{thm:SLLN_regenerative}
Let $\{X_t\}_{t=1}^{\infty}$, $X_t\in\mathcal{X}$, be a non-delayed regenerative process with associated epoch duration $\tau$, and let $f:\mathcal{X}\to\mathbb{R}$.
If $\mathbb{E}\tau<\infty$ and $\mathbb{E}[\sum_{t=1}^{\tau}|f(X_t)|]<\infty$ then:
\[
\lim_{n\to\infty}\frac{1}{n}\sum_{t=1}^{n}f(X_t)=\frac{1}{\mathbb{E}\tau}\mathbb{E}\left[\sum_{t=1}^{\tau}f(X_t)\right]
\quad\text{a.s.}
\]
\end{theorem}
This is an immediate consequence of Theorem 3.1 in~\cite[Ch.~VI]{asmussen2008applied} or of the renewal reward theorem~\cite[Prop.~7.3]{ross2014introduction}.

%We now establish the approximate optimality of this power control policy for the Bernoulli case. 
Equipped with this theorem, we now consider the process $g_t(E^t)$ obtained by our power control policy for Bernoulli energy arrivals.
We show in Appendix~\ref{subsec:power_allocation_initial_battery_state} that the initial battery level is irrelevant to the long-term average throughput (similarly to Proposition~\ref{prop:initial_battery_state}, this follows from the fact that we can always wait until the battery recharges to $\bar{B}$ before starting transmission, with a vanishing penalty to the average throughput).
We can therefore assume without loss of generality that $B_0=\bar{B}$, which necessarily implies $B_1=\bar{B}$. Hence we can equivalently assume $E_1=\bar{B}$.
Denote by $L$ the random time between two consecutive energy arrivals. Evidently, $L\sim\mathrm{Geometric}(p)$. That is,
\[
\Pr(L=k)=p(1-p)^{k-1}
\qquad,k=1,2,\ldots
\]
Observe that $g_t(E^t)$ is a non-delayed regenerative process with epoch duration $L$. 
We apply Theorem~\ref{thm:SLLN_regenerative} with $f(x)=\frac{1}{2}\log(1+x)$.
Note that $\mathbb{E}L=1/p<\infty$ and 
$\mathbb{E}[\sum_{t=1}^{L}|\tfrac{1}{2}\log(1+g_t(E^t)|]\leq\mathbb{E}[L\cdot\tfrac{1}{2}\log(1+\bar{B})]<\infty$, so the conditions of the theorem are satisfied. We obtain
\begin{align}
&\lim_{n\to\infty}\frac{1}{n}\sum_{t=1}^{n}\frac{1}{2}\log\big(1+g_t(E^t)\big)\nonumber\\*
&\qquad =\frac{1}{\mathbb{E}L}\mathbb{E}\left[\sum_{t=1}^{L}\frac{1}{2}\log\big(1+g_t(E^t)\big)\right]\quad\text{a.s.}
\label{eq:SLLN_power_control_policy}
\end{align}

We proceed to lower bound the average throughput obtained by our suggested power control policy, which is itself a lower bound to $\Conline$:
\begin{align}
%\hspace{1em}\lefteqn{\hspace{-1em}
\Conline
%}\nonumber\\*
&\geq\liminf_{n\to\infty}\frac{1}{n}\sum_{t=1}^n \mathbb{E}\left[\frac{1}{2}\log (1+g_t(E^t))\right]\nonumber\\
&\overset{\text{(i)}}{\geq}\mathbb{E}\left[\liminf_{n\to\infty}\frac{1}{n}\sum_{t=1}^{n}\frac{1}{2}\log(1+g_t(E^t))\right]\nonumber\\
&\overset{\text{(ii)}}{=}\mathbb{E}\left[\frac{1}{\mathbb{E}L}\mathbb{E}\left[\sum_{t=1}^{L}\frac{1}{2}\log(1+g_t(E^t))\right]\right]\nonumber\\
&=\frac{1}{\mathbb{E}L}\mathbb{E}\left[\sum_{t=1}^{L}\frac{1}{2}\log(1+g_t(E^t))\right]\nonumber\\
&\overset{\text{(iii)}}{=}\frac{1}{\mathbb{E}L}\mathbb{E}\left[\sum_{i=1}^{L}\frac{1}{2}\log(1+\bar{B}p(1-p)^{i-1})\right]\nonumber\\
&\overset{\text{(iv)}}{\geq}\frac{1}{\mathbb{E}L}\mathbb{E}\left[\sum_{i=1}^{L}\left[\frac{1}{2}\log(1+p\bar{B})+(i-1)\frac{1}{2}\log(1-p)\right]\right]\nonumber\\
&=\frac{1}{\mathbb{E}L}\mathbb{E}\left[L\frac{1}{2}\log(1+p\bar{B})+\frac{L(L-1)}{2}\frac{1}{2}\log(1-p)\right]\nonumber\\
&=\frac{1}{2}\log(1+p\bar{B})-\frac{1}{4}\left(\frac{\mathbb{E}[L^2]}{\mathbb{E}L}-1\right)\log\left(\frac{1}{1-p}\right)\label{eq:throughput_lower_bound_L_Bernoulli}\\
&\overset{\text{(v)}}{=}\frac{1}{2}\log(1+p\bar{B})-\frac{1}{4}\left(\frac{2-3p+p^2}{(1-p)p}-1\right)\log\left(\frac{1}{1-p}\right)\nonumber\\
&=\frac{1}{2}\log(1+p\bar{B})-\frac{1-p}{2p}\log\left(\frac{1}{1-p}\right),\label{eq:Bern}
\end{align}
where (i) is by Fatou's lemma~\cite[Thm.~1.5.4]{durrett2010probability}; (ii) is due to~\eqref{eq:SLLN_power_control_policy}; (iii) is by definition of the power control policy; (iv) is due to the inequality $\log(1+\alpha x)\geq\log(1+x)+\log\alpha$ for $0<\alpha\leq 1$;
and (v) is because $L\sim\mathrm{Geometric}(p)$.

%The gap from the upper bound is largest when $p\to0$, in which case it is given by $\frac{1}{2\ln 2}\approx0.7213$ bits.
The second term in the above expression achieves its maximum when $p\to0$, in which case it is given by $\frac{1}{2\ln 2}\approx 0.72$. We conclude that for Bernoulli energy arrivals:
\begin{equation}
\label{eq:Bernoulli_lower_bound}
\Conline\geq\frac{1}{2}\log(1+\mu)-\frac{1}{2\ln 2},
\end{equation}
where $\mu=\mathbb{E}[\min\{E_t,\bar{B}\}]=p\bar{B}$ is the average energy arrival rate of the Bernoulli process.

\subsubsection{General i.i.d Energy Harvesting Processes}
\label{subsubsec:general_energy_arrivals}
We next turn to developing approximately optimal online power control policies for general i.i.d. energy harvesting processes. A simple extension of the Bernoulli power control policy presented in the previous section to general i.i.d. processes was proposed  in~\cite{DongOzgur2014}. \cite{DongOzgur2014} also showed, via providing examples, that this extension can achieve the long-term average throughput within a constant gap for some i.i.d. processes, but it would fail to do so for some others. \cite{DongOzgur2014} however did not explicitly characterize the distributions for which this extension achieves the long-term average throughput within a constant gap. Below, we first overview the policy proposed in \cite{DongOzgur2014}, calling it the \emph{binary quantization} policy and then show that its gap to optimality depends on $\mu/\bar{B}$. In particular, the gap becomes unbounded when $\mu/\bar{B}\to 0$. We then construct a new online power control policy, which we call the \emph{generalized Bernoulli} policy which approximately achieves the long-term average throughput when  $\mu/\bar{B}$ is small. Considering 
these two policies in the large and small $\mu/\bar{B}$ regimes respectively proves Theorem~\ref{thm:onlinePC}.
\bigbreak
\paragraph*{The Binary Quantization Policy} Let $E_t$ be an arbitrary i.i.d. energy arrival process. Note that as before, given a battery size $\bar{B}$ we can concentrate on the equivalent process $\tilde{E}_t=\min\{E_t,\bar{B}\}$  with mean $\mu=\mathbb{E}[\tilde{E}_t]$. Let the complementary cumulative distribution function (ccdf) $\bar{F}(x)=\Pr\{\tilde{E}_t\geq x\}$, $x\in[0,\bar{B}]$.
Consider the following policy: fix an energy level $x$. Denote by $q'$ the probability of observing an energy arrival at least $x$, i.e. $q'\triangleq\bar{F}(x)$.
Then, apply the exponentially decreasing strategy described above as if $\tilde{E}_t$ is i.i.d. Bernoulli with levels $\{0,x\}$ and probability $q'$.
Effectively, when ${\tilde{E}_t\geq x}$, we treat the incoming energy as a packet of size $x$ and ignore the remaining energy. Alternatively, when $\tilde{E}_t<x$, we ignore the incoming energy completely and assume no new energy has arrived. This  of course is an admissible policy, but may be highly suboptimal as it ignores part of the incoming energy. Nevertheless, we will see below that, for a certain class of energy arrival processes, it differs from the upper bound only by a constant gap.

To begin, we apply~\eqref{eq:Bernoulli_lower_bound} to lower bound the rate obtained by this strategy:
$\Conline \geq \frac{1}{2}\log\left(1+x\bar{F}(x)\right) - \frac{1}{2\ln 2}$.
The tightest lower bound will be obtained by maximizing over $x\in[0,\bar{B}]$:
\[
\Conline \geq \frac{1}{2}\log\left(1+
\max_{x\in[0,\bar{B}]}\{x\bar{F}(x)\}\right)
 - \frac{1}{2\ln 2}.
\]
We will now upper bound the gap of this policy by providing a lower bound of the form 
\[
\max_{x\in[0,\bar{B}]}x\bar{F}(x) \geq c\mu,
\]
for $c\in[0,1]$. To this end, we introduce the Lambert $W$ function, which is defined as the solution to $z=W(z)e^{W(z)}$.
This function is double-valued on $(-1/e,0)$ -- that is, for $z$ in this interval there are two possible solutions for the above transcendental equation.
Specifically, we are interested in the lower branch, denoted by $W_{-1}(z)$, which is defined for $z\in[-1/e,0)$. This function is strictly decreasing, and it decreases from $W_{-1}(-e^{-1})=-1$ to $\lim_{z\nearrow 0} W_{-1}(z)=-\infty$. We now show that 
\begin{equation}
\label{eq:cstar_def}
c^*= c^*\left(\frac{\mu}{\bar{B}}\right) \triangleq -\frac{1}{W_{-1}\left(-\frac{\mu}{\bar{B}}e^{-1}\right)}
\end{equation}
is one such $c$. 
Note that since $\tilde{E}_t$ is of bounded support $[0,\bar{B}]$, we always have $0\leq\mu/\bar{B}\leq 1$, so the above expression is well-defined.
Also note that the definition of $c^*$ implies
\[
-\frac{\mu}{\bar{B}}e^{-1}=-\frac{1}{c^*}e^{-1/c^*},
\]
\begin{equation}
1-c^*=c^*\ln\left(\frac{\bar{B}}{\mu c^*}\right).
\label{eq:c_equation}
\end{equation}
Now, suppose $x\bar{F}(x)<c^*\mu$, $\forall x\in[0,\bar{B}]$. We have:
\begin{align*}
\mu
&=\int_0^{\bar{B}}\bar{F}(x)dx\\
&\leq \int_0^{c^*\mu}dx + \int_{c^*\mu}^{\bar{B}}\bar{F}(x)dx\\
&< c^*\mu + \int_{c^*\mu}^{\bar{B}}\frac{c^*\mu}{x}dx\\
&= c^*\mu + c^*\mu\ln\left(\frac{\bar{B}}{c^*\mu}\right)
\end{align*}
This yields $c^* + c^*\ln\left(\frac{\bar{B}}{c^*\mu}\right) > 1$, which contradicts the definition of $c^*$~\eqref{eq:c_equation}. 

Finally, we obtain:
\begin{align}
\Conline&\geq\frac{1}{2}\log(1+c^*\mu)-\frac{1}{2\ln2}\nonumber\\
&\geq\frac{1}{2}\log(1+\mu)-\frac{1}{2}\log\left(\frac{e}{c^*}\right),
\label{eq:Conline_lower_first}
\end{align}
where the second step is due to the inequality $\log(1+\alpha x)\geq\log(1+x)+\log\alpha$ for $0<\alpha\leq 1$, and $c^*(\mu/\bar{B})$ is defined in~\eqref{eq:cstar_def}.

%Notice that for this scheme, $H(g^n)\leq H(\tilde{E}^n) = nH(p^*)$, since $g^n$ is a function of $\tilde{E}^n$, where $p^* = 1-F(x^*)$ for some $x^*$ such that $x^*(1-F(x^*))\geq c^*\mu$. As a corollary, we have:
%\begin{align*}
%&\lim_{n\to\infty}\max_{g^n\in\Gonline(\bar{B})} \left\{\T(g^n)-\tfrac{1}{n}H(g^n)\right\} \\
%&\geq \frac{1}{2}\log(1+\mu) - \log\frac{1}{c^*} - \frac{1-p^*}{2p^*}\log\left(\frac{1}{1-p^*}\right) - H(p^*)\\
%&\geq \frac{1}{2}\log(1+\mu) - \log\frac{1}{c^*} - 1.5242
%\end{align*}

Note that the gap of the policy to the upper bound ${\frac{1}{2}\log(1+\mu)}$ depends on the parameters of the problem $\mu$ and $\bar{B}$ through $c^*$.
One observes that $c^*\to 0$ as $\mu/\bar{B}\to 0$, making the gap unbounded. This suggests that the \emph{binary quantization} policy does not work well for small $\mu/\bar{B}$. For such distributions $x\bar{F}(x)$ can be much smaller than $\mu$, which implies that a significant amount of incoming energy is discarded by the policy. This indeed  is the case for the counterexample presented in~\cite[Section VI.C]{DongOzgur2014}. In the sequel, we will present a more interesting generalization of the Bernoulli policy, which  achieves a finite gap for the range of small $\mu/\bar{B}$. Choosing the appropriate policy out of the two, depending on the value of $\mu$, will provide a bounded gap for all values of $\mu\in[0,\bar{B}]$.
\bigbreak
\paragraph*{The Generalized Bernoulli Policy}
%As discussed in the last section, a good allocation policy should retain the exponentially decaying structure and not waste too much energy, while still satisfying the power constraints of the system. Our suggestion is as follows. 
Let $q\triangleq\mu/\bar{B}$, where recall that $\mu=\mathbb{E}[\tilde{E}_t]$ and $\tilde{E}_t=\min\{E_t,\bar{B}\}$. Note that ${\mu\in(0,\bar{B}]}$ so $q\in(0,1]$.
Consider the following energy allocation policy: 
\[
g_t=\bar{B}q(1-q)^{t-s_t}=\mu(1-q)^{t-s_t},
\]
where 
\[
s_t=s_t(\tilde{E}^t,g^{t-1})=\{\sup\ \tau\leq t:\ B_{\tau}=\bar{B}\}.
\]
That is, $s_t$ is the last time the battery was completely full.
This is clearly an admissible online power control policy, since, as before, even if the battery never gets recharged, the total energy used will not exceed $\bar{B}$.

\begin{figure}
\centering
\begin{tikzpicture}
	\tikzstyle{every node}=[font=\scriptsize];
	
	\foreach \x/\g/\i in {0/1/1, 0.5/0.5/2, 1/0.25/3, 1.5/0.125/4, 2/1/1, 2.5/0.5/2, 3/0.25/3, 3.5/1/1, 4/0.5/2}
	{
		\FPeval{\mycolor}{round(100*\g,0)};
		\filldraw[fill=gray!\mycolor!white,draw=black] (\x,0) -- (\x+.5,0) -- (\x+.5,\g) -- (\x,\g) -- cycle;
		\node[above] at (\x+.25,\g) {$g_{\i}$};
	}
	
	\node at (-.5,.7) {$g_t$};	
	
	\def \h {2};
	
	\foreach \x/\b/\e/\i/\mygreen/\myyellow in {0/2/0.2/1/100/100, 0.5/1.4/0.4/2/0/80, 1/1.5/0.6/3/0/100, 1.5/1.65/0.4/4/30/100, 2/2/0.8/5/100/100, 2.5/1/0/6/0/0, 3/1.7/1.2/7/40/100, 3.5/2/0.6/8/100/100, 4/1.2/0.2/9/0/40}
	{
		%\FPeval{\mycolor}{round(65*\b-30,0)};
		\FPeval{\mycolor}{round(50*\b,0)};
		\filldraw[fill=gray!\mycolor!white,draw=black] (\x,\h) -- (\x+.5,\h) -- (\x+.5,\h+\b) -- (\x,\h+\b) -- cycle;
		\draw[->] (\x+.25,\h+2.8) node[above] {$\e$} -- (\x+.25,\h+2.4);
	}
	
	\node[above] at (-.5,\h+2.8) {$E_t$};
	\node at (-.5,\h+1) {$B_t$};

\end{tikzpicture}
\caption{The generalized Bernoulli power control policy for general i.i.d. energy arrivals.}
\label{fig:genBernoulli}
\end{figure}
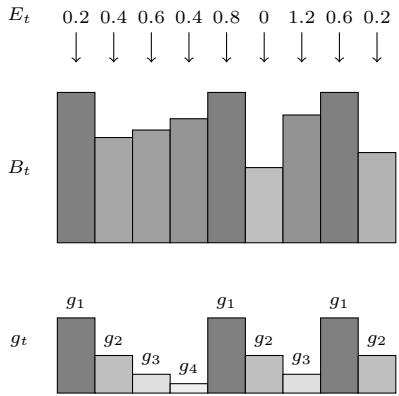

Notice the similarity between the scheme in the previous section and the \emph{generalized Bernoulli} policy: we transmit using an exponentially decreasing power allocation policy, and ``restart'' whenever the battery recharges completely. In the Bernoulli case, the event of battery recharge was an i.i.d. process depending solely on the energy arrivals. Here, it depends both on the energy arrivals $\tilde{E}_t$ (which have an arbitrary distribution) and on the sequence of powers $g_t$. See Figure~\ref{fig:genBernoulli}.

By construction, the sequence $g_t$ is a regenerative process; on the event of battery recharge, the power control policy ``restarts'', and by the i.i.d. nature of the energy arrivals, the portions of the process between consecutive battery recharges are independent and identically distributed.
This lends to a similar analysis as in the Bernoulli case, and we can lower bound the average throughput following the steps in the previous section to obtain an analog of~\eqref{eq:throughput_lower_bound_L_Bernoulli}:
\[
\Conline \geq \frac{1}{2}\log(1+\mu)-\frac{1}{4}\left(\frac{\mathbb{E}[L^2]}{\mathbb{E}L}-1\right)\log\left(\frac{1}{1-q}\right)
\]
where $L$ is the time between consecutive battery recharges.
However, $L$ is no longer geometrically distributed, but has some distribution that depends on $\tilde{E}_t$ and $g_t$.
Hence, to have applied Theorem~\ref{thm:SLLN_regenerative} in this case to obtain the above lower bound, we need to verify that $L$ has finite expectation; this is done in Appendix~\ref{sec:L_finite_expectation}.

We wish to upper bound the gap from $\frac{1}{2}\log(1+\mu)$:
\[
\mathrm{gap}=\frac{1}{4}\left(\frac{\mathbb{E}[L^2]}{\mathbb{E}L}-1\right)\log\left(\frac{1}{1-q}\right).
\]
To this matter, consider the first epoch (or cycle) of the regenerative process $g_t(\tilde{E}^t)$.
The length of the epoch $L$ can be defined as the first time for which $B_{L+1}=\bar{B}$ (i.e. the battery at the beginning of the following epoch is fully charged). This implies that $B_t<\bar{B}$ for $t=2,\ldots,L$ (recall $B_1=\bar{B}$), therefore
$B_t=B_{t-1}-g_{t-1}+\tilde{E}_t$.
Substituting $g_t=\bar{B}q(1-q)^{t-1}$, we obtain:
\[
B_t=\bar{B}(1-q)^{t-1}+\sum_{i=2}^{t}\tilde{E}_i.
\]
Denote $S_t\triangleq\sum_{i=2}^{t+1}\tilde{E}_i$.
Hence, $L$ is the first time for which
$B_{L}-g_L+\tilde{E}_{L+1}\geq\bar{B}$, or:
\[
L=\{\inf\ t:\ S_t\geq\bar{B}[1-(1-q)^t]\}.
\]
This is a stopping time adapted to $\tilde{E}_{t+1}$, so by Wald's first and second identities (cf. Theorems~4.1.5 and~4.1.6 in~\cite{durrett2010probability}):
\begin{align}
\mathbb{E}[S_L]&=\mathbb{E}L\cdot\mu,\label{eq:Wald1}\\
\mathbb{E}[(S_L-\mu L)^2]&=\sigma^2\mathbb{E}L,\label{eq:Wald2}
\end{align}
where $\sigma^2\triangleq\mathrm{Var}(\tilde{E}_t)$.
We obtain:
\begin{equation}
\mathbb{E}[L^2]=\frac{1}{\mu^2}\big(\sigma^2\mathbb{E}L +2\mu\mathbb{E}[LS_L]-\mathbb{E}[S_L^2]\big).
\label{eq:Lsquared_lower_bound}
\end{equation}
Next, by definition $S_{L-1}<\bar{B}[1-(1-q)^{L-1}]$, and since $\tilde{E}_t\leq\bar{B}$,
\begin{equation*}
S_L=S_{L-1}+E_{L+1}<\bar{B}[1-(1-q)^{L-1}]+\bar{B}
\leq 2\bar{B}.
\end{equation*}
Moreover, 
\begin{equation*}
\frac{\mathbb{E}[S_L^2]}{\mathbb{E}L}
\overset{\text{(i)}}{\geq}\frac{(\mathbb{E}[S_L])^2}{\mathbb{E}L} 
\overset{\text{(ii)}}{=}\mu^2\cdot\mathbb{E}L
\overset{\text{(iii)}}{\geq}\mu^2,
\end{equation*}
where (i) is true for any r.v.; (ii) is due to~\eqref{eq:Wald1}; and (iii) is because $L\geq1$.
Plugging these two inequalities in~\eqref{eq:Lsquared_lower_bound} yields:
\begin{align*}
\frac{\mathbb{E}[L^2]}{\mathbb{E}L}
&=\frac{1}{\mu^2}\left(\sigma^2+2\mu\frac{\mathbb{E}[LS_L]}{\mathbb{E}L}-\frac{\mathbb{E}[S_L^2]}{\mathbb{E}L}\right)\\
&\leq\frac{1}{\mu^2}(\sigma^2+4\mu \bar{B}-\mu^2).
\end{align*}
Next, observe that since $0\leq \tilde{E}_t\leq\bar{B}$, then $\tilde{E}_t^2\leq \tilde{E}_t\bar{B}$, and therefore $\mathbb{E}[\tilde{E}_t^2]\leq\mu\bar{B}$, or $\sigma^2\leq\mu(\bar{B}-\mu)$ (This result, in a more general form, is called the Bhatia-Davis inequality~\cite{BhatiaDavis2000}).
We conclude that
\begin{equation*}
\frac{\mathbb{E}[L^2]}{\mathbb{E}L}\leq 5\frac{\bar{B}}{\mu}-2 =\frac{5}{q}-2,
\end{equation*}
which yields
\begin{equation*}
\mathrm{gap} \leq \frac{5-3q}{4q}\log\left(\frac{1}{1-q}\right),
\end{equation*}
\begin{equation}
\Conline\geq\frac{1}{2}\log(1+\mu)-\frac{5-3q}{4q}\log\left(\frac{1}{1-q}\right).
\label{eq:Conline_lower_second}
\end{equation}
Observe that the gap is finite for $q\to0$.

Finally, combining~\eqref{eq:Conline_lower_first} and~\eqref{eq:Conline_lower_second}, we have:
\begin{align*}
\hspace{1em}\lefteqn{\hspace{-1em}
\frac{1}{2}\log(1+\mu) - \Conline
}\\*
&\leq \max_{0\leq q\leq 1} \min \left\{\frac{1}{2}\log\frac{e}{c^*(q)}, \frac{5-3q}{4q}\log\left(\frac{1}{1-q}\right)\right\}\\
&\leq 1.8034,
\end{align*}
which completes the proof of Theorem~\ref{thm:onlinePC}.

\subsection{Entropy of the Power Control Policies}\label{sec:entropyrate}
In the light of \eqref{eq:causal_lower_bound} in Theorem~\ref{thm:bounds}, we care not only about the long-term average throughput achieved by a certain online power control policy, but also its entropy per symbol $\tfrac{1}{n}H(g^n(E^n))$, which determines the gap between the information-theoretic capacity and the long-term average throughput in the case where the receiver does not have energy arrival information. We next show that the per-symbol entropies of the power control policies $g^n$ we developed in the previous section can be bounded by $1$ bit/channel use, and in this manner prove Proposition~\ref{prop:prop1}. This is due to the structure of these processes and their regenerative nature. All the randomness in the processes $g^n(E^n)$ is contained in the epoch start times; knowing where the epochs begin is sufficient to generate the corresponding $g^n(E^n)$ for all the policies discussed in the previous section.

We start with the policy for Bernoulli energy arrivals discussed in Section~\ref{subsubsec:Bernoulli}. Observe that
\[
\frac{1}{n}H(g^n(E^n))\leq \frac{1}{n}H(E^n)=H(E_t)=H_2(p),
\]
which combined with \eqref{eq:Bern} implies
\begin{align*}
\hspace{2em}\lefteqn{\hspace{-2em}
\liminf_{n\to\infty}\max_{g^n\in\Gonline(\bar{B})}
\{\T(g^n)-\tfrac{1}{n}H(g^n(E^n))\}
}\\*
&\geq \frac{1}{2}\log(1+p\bar{B})-\frac{1-p}{2p}\log\left(\frac{1}{1-p}\right)-H_2(p).
\end{align*}
The expression $\frac{1-p}{2p}\log(1-p)+H_2(p)$ is upper bounded by 1.5242 for all $p\in[0,1]$, hence we have
\begin{align}
\lefteqn{
\liminf_{n\to\infty}\max_{g^n\in\Gonline(\bar{B})}
\{\T(g^n)-\tfrac{1}{n}H(g^n(E^n))\}
}\nonumber\\*
&
\hspace{10em}
\geq \frac{1}{2}\log(1+\mu)-1.5242,
\label{eq:Bernoulli_lower_bound_noCSIR}
\end{align}
where recall that $\mu=p\bar{B}$ for the Bernoulli process.

Next, we bound the entropy of the \emph{binary quantization} policy. Define $E'_t=1\{E_t\geq x\}$ and observe that for this policy
\[
\tfrac{1}{n}H(g^n(E^n))=\tfrac{1}{n}H(g^n(E'^n))
\leq \tfrac{1}{n}H(E'^n)
 =H_2(\bar{F}(x)).
\]
Then, similarly to~\eqref{eq:Bernoulli_lower_bound_noCSIR}, we have:
\begin{align}
\hspace{3em}\lefteqn{\hspace{-3em}
\liminf_{n\to\infty}\max_{g^n\in\Gonline(\bar{B})}
\{\T(g^n)-\tfrac{1}{n}H(g^n(E^n))\}
}\nonumber\\*
&\geq \frac{1}{2}\log(1+c^*\mu)
	-H_2(q')-\frac{1-q'}{2q'}\log\left(\frac{1}{1-q'}\right)\nonumber\\
&\geq \frac{1}{2}\log(1+\mu)-\frac{1}{2}\log\frac{1}{c^*}-1.5242,
\label{eq:rate-H_lower_first}
\end{align}
where $q'=\bar{F}(x)$ as defined in Section~\ref{subsubsec:general_energy_arrivals}.

Finally, we focus on the \emph{generalized Bernoulli} policy. Define the indicator process $F_t = 1(B_t=\bar{B})$.
%Since the energy harvesting profile is known, $g^n$ is a function of $F^n$.
By construction, $F^t$ is enough to determine $g_t(E^t)$, hence
\[
\tfrac{1}{n}H(g^n(E^n))
=\tfrac{1}{n}H(g^n(F^n))
\leq\tfrac{1}{n}H(F^n)
\leq 1.
\]
%Therefore, $H(g^n)\leq H(F^n) \leq n$. As a corollary, we obtain:
We therefore have from~\eqref{eq:Conline_lower_second}:
\begin{align}
&
\liminf_{n\to\infty}\max_{g^n\in\Gonline(\bar{B})} \left\{\T(g^n)-\tfrac{1}{n}H(g^n(E^n))\right\}
\nonumber\\*
&
\hspace{2em}
\geq \frac{1}{2}\log(1+\mu) - \frac{5-3q}{4q}\log\left(\frac{1}{1-q}\right) - 1.
\label{eq:rate-H_lower_second}
\end{align}

Now, combining~\eqref{eq:rate-H_lower_first} and~\eqref{eq:rate-H_lower_second}:
\begin{align*}
\lefteqn{
\frac{1}{2}\log(1+\mu) - \liminf_{n\to\infty}\max_{g^n\in\Gonline(\bar{B})} \left\{\T(g^n)-\tfrac{1}{n}H(g^n(E^n))
\right\}
}\\*
&\leq \max_{0\leq q\leq 1} \min \left\{\frac{1}{2}\log\frac{1}{c^*(q)} + 1.5242, \frac{5-3q}{4q}\log\left(\frac{1}{1-q}\right) + 1\right\}\\
&\leq 2.8034
\end{align*}
which, along with~\eqref{eq:power_allocation_upper_bound}, gives the result of Proposition~\ref{prop:prop1}.

%---------------------------%
\section{Connection between Capacity and Throughput: Proof of Theorem~\ref{thm:bounds}}
%---------------------------%
\label{sec:capacity_bounds}

%We prove Theorem~\ref{thm:bounds}.% for $\CcausalTx$ and $\CcausalTxRx$.
%The proof for $\CnoncausalTxRx$ follows the same lines, and is brought in Appendix~\ref{subsec:noncausal_bounds_proof}.
In this section we upper and lower bound the capacity of the energy harvesting channel, given by the expressions in Theorem~\ref{thm:capacity}, using the solution to the power control optimization problem.

%---------------------------%
\subsection{Upper Bounds}
%---------------------------%
\label{subsec:upper_bounds}

We start with $\CcausalTxRx$.
Note that since $\CcausalTx\leq\CcausalTxRx$, we will obtain both~\eqref{eq:causal_upper_bound} and the upper bound for~\eqref{eq:causalRx_bounds}.
%It is shown in Appendix~\ref{subsec:alternative_constraints} 
Through some algebraic manipulations, we show in Appendix~\ref{sec:alternative_constraints} that $\mathcal{Q}_n(b)$, defined in~\eqref{eq:Qn_definition}, can be written as
\begin{align*}
	\mathcal{Q}_n(b)=\Big\{&P_{X^n\|E^n}\text{ s.t. $\forall e^n\in\mathcal{E}^n$, a.s. for 
		$t=1,\ldots,n$:}\\*
	&\sum_{j=i}^{t}X_j^2\leq\bar{B}+\sum_{j=i+1}^{t}e_j
	\quad,i=1,\ldots,t,\\
	&\sum_{j=1}^{t}X_j^2\leq b+\sum_{j=1}^{t}e_j\Big\}.
\end{align*}
Next, we define another set of probability distributions by relaxing the a.s. constraints to hold in expectation:
\begin{align*}
	\mathcal{Q}^\ast_n(b)=\Big\{&P_{X^n\|E^n}\text{ s.t. for $t=1,\ldots,n$ and $\forall e^n\in\mathcal{E}^n$:}\\*
	&\sum_{j=i}^{t}\mathbb{E}\big[X_j^2|E^j=e^j\big]\leq\bar{B}+\sum_{j=i+1}^{t}e_j\\*
	&\hspace{12em} ,i=1,\ldots,t,\\*
	&\sum_{j=1}^{t}\mathbb{E}\big[X_j^2|E^j=e^j\big]\leq b+\sum_{j=1}^{t}e_j\Big\}.
\end{align*}
Observe that $\mathcal{Q}_n(b)\subseteq\mathcal{Q}_n^\ast(b)$.

By the same arguments as before,
we can write \eqref{eq:G_definition} for the set of online policies as:
\begin{align}
	\Gonline(b)=\Big\{&g^n=(g_1,\ldots,g_n),\ g_t:\mathcal{E}^t\to\mathbb{R}_+,\nonumber\\*
	& \text{ s.t. }\forall e^n\in\mathcal{E}^n:\nonumber\\*
	&\sum_{j=i}^{t}g_j(e^j)\leq\bar{B}+\sum_{j=i+1}^{t}e_j
		\quad,i=1,\ldots,t,\nonumber\\*
	&\sum_{j=1}^{t}g_j(e^j)\leq b+\sum_{j=1}^{t}e_j
		\quad,t=1,\ldots,n\Big\}.
	\label{eq:G_alt_def}
\end{align}

Next, we upper bound the mutual information in~\eqref{eq:causalRx_capacity} as follows:
\begin{align*}
I(X^n;Y^n|E^n)
&=h(Y^n|E^n)-h(Y^n|X^n,E^n)\\
&\leq\sum_{t=1}^{n}[h(Y_t|E^t)-h(Y_t|X_t,E^t)]\\
&=\sum_{t=1}^{n}I(X_t;Y_t|E^t),
\end{align*}
where the inequality is due to the memorylessness of the channel.
Applying this to~\eqref{eq:causalRx_capacity} gives:
\begin{align}
	\CcausalTxRx&=\lim_{n\to\infty}\frac{1}{n}
		\sup_{P_{X^n\|E^n}\in\mathcal{Q}_n(b)}
		I(X^n;Y^n|E^n)\nonumber\\*
	&\leq\liminf_{n\to\infty}\frac{1}{n}
		\sup_{P_{X^n\|E^n}\in\mathcal{Q}_n(b)}
		\sum_{t=1}^{n}I(X_t;Y_t|E^t)\nonumber\\
	&\leq
		\liminf_{n\to\infty}\frac{1}{n}
		\sup_{P_{X^n\|E^n}\in\mathcal{Q}^\ast_n(b)}
		\sum_{t=1}^{n}I(X_t;Y_t|E^t),
		\label{eq:CcausalTxRx_sum_scalar_channels}
\end{align}
where the last inequality is because $\mathcal{Q}_n(b)\subseteq\mathcal{Q}_n^\ast(b)$.
Note that we always have
\[
I(X_t;Y_t|E^t=e^t)
\leq\frac{1}{2}\log\left(1+\mathbb{E}
	\big[X_t^2|E^t=e^t\big]\right),
\]
since the mutual information of the scalar AWGN channel is always maximized by a Gaussian input distribution.
Taking expectation, we obtain:
\[
I(X_t;Y_t|E^t)\leq\mathbb{E}\left[\frac{1}{2}
		\log\big(1+\mathbb{E}[X_t^2|E^t]\big)
		\right].
\]
Plugging this into~\eqref{eq:CcausalTxRx_sum_scalar_channels} yields:
\[
\CcausalTxRx\leq
\liminf_{n\to\infty}\frac{1}{n}
\sup_{P_{X^n\|E^n}\in\mathcal{Q}_n^*(b)}
\mathbb{E}\left[\sum_{t=1}^{n}
\frac{1}{2}\log\big(1+\mathbb{E}[X_t^2|E^t]\big)
\right].
\]

For a fixed $P_{X^n\|E^n}\in\mathcal{Q}_n^\ast(b)$ and for each $e^n\in\mathcal{E}^n$, denote $g_t(e^t)=\mathbb{E}[X_t^2|E^t=e^t]$, $t=1,\ldots,n$.
Then $g^n\in\Gonline(b)$ as given by~\eqref{eq:G_alt_def}.
Therefore,
\begin{align*}
	\CcausalTxRx&\leq\liminf_{n\to\infty}\frac{1}{n}
		\max_{g^n\in\Gonline(b)}
		\mathbb{E}\left[\sum_{t=1}^{n}
		\frac{1}{2}\log(1+g_t(E^t))\right]\\*
	&=\liminf_{n\to\infty}
		\max_{g^n\in\Gonline(b)}
		\T(g^n)\\*
	&=\Conline.
\end{align*}
This gives~\eqref{eq:causal_upper_bound} and the upper bound in~\eqref{eq:causalRx_bounds}.
The derivation of the upper bound in~\eqref{eq:noncausal_bounds} is similar, and is shown in Appendix~\ref{sec:noncausal_bounds_proof}.
%Applying similar arguments for $\CnoncausalTxRx$, we can show the RHS of~\eqref{eq:noncausal_bounds}.

%---------------------------%
\subsection{Lower Bounds}
%---------------------------%
\label{subsec:lower_bounds}

We derive here the lower bounds in~\eqref{eq:causal_bounds} and~\eqref{eq:causalRx_bounds}. The derivation of the lower bound in~\eqref{eq:noncausal_bounds} is similar, and is deferred to Appendix~\ref{sec:noncausal_bounds_proof}.

\subsubsection{Energy Arrival Information at the Transmitter and the Receiver}
We start with $\CcausalTxRx$.
Fix $g^n\in\Gonline(b)$.
$g^n$ will determine an energy allocation policy for transmission, and at time $t=1,\ldots,n$ we transmit a symbol with a peak power constraint of $g_t(E^t)$.
%that maximizes the instantaneous rate with peak power constraint $g_t(E^t)$.
%Let $p^\star(x;\mathcal{E})$ be the distribution that maximizes $I(X;Y)$ out of all distributions with support $X^2\leq\mathcal{E}$.
%More precisely, each symbol is drawn from the distribution
%\begin{equation}
%\label{eq:smith_capacity_def}
%	P^{\star[S]}_{X}\triangleq
%	\argmax_{P_X:\ X^2\leq S\text{ a.s.}}
%	I(X;X+N),
%\end{equation}
%where $N\sim\mathcal{N}(0,1)$.
%That is, $P_X^{\star[S]}$ is the capacity achieving input distribution for the amplitude constrained scalar Gaussian channel.
More precisely, for every $S\in[0,\bar{B}]$, fix a distribution $P_X^{[S]}$ with support $[-\sqrt{S},\sqrt{S}]$.
We construct an input distribution of the form
\[
P_{X^n\|E^n}(x^n\|e^n)
=\prod_{t=1}^{n}P_{X_t|E^t}(x_t|e^t),
\]
where 
%$P_{X_t|E^t}(x_t|e^t)=P^{\star[g_t(e^t)]}_X(x_t)$.
$P_{X_t|E^t}(x_t|e^t)=P^{[g_t(e^t)]}_X(x_t)$.
Since $X_t^2\leq g_t(E^t)$ and $g^n$ is an admissible online power control policy, the energy constraints are satisfied completely.
This is clearly suboptimal, since most likely for some $t$, $X_t^2<g_t(E^t)$, therefore energy will be wasted.
Still, $P_{X^n\|E^n}\in\mathcal{Q}_n(b)$,
and thus we can obtain a lower bound by computing the mutual information in~\eqref{eq:causalRx_capacity} for $P_{X^n\|E^n}$.
Note that the $X_t$'s are independent given $E^n$. In fact, we have:
\[
P_{X^n,Y^n|E^n}(x^n,y^n|e^n)
=\prod_{t=1}^{n}P_{X_t|E^t}(x_t|e^t)P_{Y|X}(y_t|x_t),
\]
therefore,
\begin{align*}
I(X^n;Y^n|E^n)
&=\sum_{t=1}^{n}I(X_t;Y_t|E^n)\\
&=\sum_{t=1}^{n}I(X_t;Y_t|E^t)\\
&=\sum_{e^n}P_{E^n}(e^n)\sum_{t=1}^{n}
	I(X_t;Y_t|E^t=e^t).
\end{align*}
%\begin{align}
%	\CcausalTxRx
%	&\geq\lim_{n\to\infty}\frac{1}{n}
%		I(X^n;Y^n|E^n)\nonumber\\
%	&=\lim_{n\to\infty}\frac{1}{n}
%		\sum_{e^n}P_{E^n}(e^n)
%		I_{e^n}(X^n;Y^n)\nonumber\\
%	&=\lim_{n\to\infty}\frac{1}{n}\sum_{e^n}P_{E^n}(e^n)
%		\sum_{t=1}^{n}I_{e^t}(X_t;Y_t)\nonumber
%%\\
%%	&=\lim_{n\to\infty}\frac{1}{n}\sum_{e^n}P_{E^n}(e^n)
%%		\sum_{t=1}^{n}\max_{X^2\leq g_t(e^t)}I(X;X+N).
%%		\label{eq:lower_bound_smith}
%\end{align}
%Observe that $I_{e^t}(X_t;Y_t)$ is in fact $I(X;X+N)$ where ${X\sim P_X^{[g_t(e^t)]}}$ and $N\sim\mathcal{N}(0,1)$ independent of $X$.
Observe that $I(X_t;Y_t|E^t=e^t)$ is in fact the rate obtained for a scalar AWGN channel when the input distribution is~$P_X^{[g_t(e^t)]}$.
We can therefore maximize over all such input distributions to obtain the following lower bound for~\eqref{eq:causalRx_capacity}:
\begin{align}
\CcausalTxRx
&\geq\limsup_{n\to\infty}\frac{1}{n}
\sum_{e^n}P_{E^n}(e^n)
\sum_{t=1}^{n}\CSmith(g_t(e^t))\nonumber\\
&=\limsup_{n\to\infty}\frac{1}{n}\mathbb{E}
\left[\sum_{t=1}^{n}\CSmith(g_t(E^t))\right],
\label{eq:lower_bound_smith}
\end{align}
where
\[
\CSmith(S)\triangleq\max_{X^2\leq S}I(X;X+N),
\]
for $N\sim\mathcal{N}(0,1)$ independent of $X$.
This is the capacity of the amplitude constrained scalar Gaussian channel, which was found in~\cite{Smith1971} (hence the notation $\CSmith$).
Unfortunately it is not tractable, however,
%However, it is not tractable.
%Nevertheless, 
%as done in~\cite{DongOzgur2014,OzarowWyner1990},
it can be lower bounded using the following lemma.
\begin{lemma}
\label{lemma:EPI_lower_bound}
The capacity of the amplitude constrained scalar AWGN channel with noise variance 1 can be lower bounded as follows:
\begin{equation}
	\label{eq:ozarow_wyner_lower_bound}
	%\max_{X^2\leq S}I(X;X+N)
	\CSmith(S)
	\geq\frac{1}{2}\log(1+S)
	-\frac{1}{2}\log\left(\frac{\pi e}{2}\right).
\end{equation}
\end{lemma}
This is the same as Lemma~1 in~\cite{DongFarniaOzgur2015}, the proof of which relies on results from~\cite{OzarowWyner1990}.
For completeness, we bring here a simple proof using the entropy power inequality.
\begin{proof}
Let $X$ be uniform on the interval $[-\sqrt{S},\sqrt{S}]$. Then, using the entropy power inequality:
\begin{align}
I(X;X+N)&=h(X+N)-h(N)\nonumber\\*
&\geq\frac{1}{2}\log\big(2^{2h(X)}+2^{2h(N)}\big)-h(N)\nonumber\\
&=\frac{1}{2}\log(4S+2\pi e)-\frac{1}{2}\log(2\pi e)\nonumber\\
&=\frac{1}{2}\log\left(1+\frac{2S}{\pi e}\right)
	\label{eq:EPI_lower_bound}\\
&\geq\frac{1}{2}\log(1+S)-\frac{1}{2}\log\left(\frac{\pi e}{2}\right).\nonumber
\end{align}
\end{proof}
Plugging~\eqref{eq:ozarow_wyner_lower_bound} into~\eqref{eq:lower_bound_smith}:
\begin{align}
	\CcausalTxRx
	&\geq\limsup_{n\to\infty}\frac{1}{n}\mathbb{E}
		\left[\sum_{t=1}^{n}\frac{1}{2}
		\log(1+g_t(E^t))\right]
		-\frac{1}{2}\log\left(\frac{\pi e}{2}\right)
		\nonumber\\
	&\geq\liminf_{n\to\infty}\T(g^n)
		-\frac{1}{2}\log\left(\frac{\pi e}{2}\right).
		\label{eq:causalTxRx_online_lower_bound}
\end{align}
Since this is true for any $g^n\in\Gonline(b)$, we can take the maximum to obtain:
\begin{align*}
%	\CcausalTxRx&\geq\lim_{n\to\infty}\frac{1}{n}
%		\max_{g^n\in\Gonline(b)}
%		\mathbb{E}\left[
%		\sum_{t=1}^{n}\frac{1}{2}\log(1+g_t(E^t))
%		\right]\\*
%		&\hspace{5.5cm}
%		-\frac{1}{2}\log\left(\frac{\pi e}{2}\right)\\
	\CcausalTxRx&\geq\liminf_{n\to\infty}
		\max_{g^n\in\Gonline(b)}
		\T(g^n)
		-\frac{1}{2}\log\left(\frac{\pi e}{2}\right)\\*
	&=\Conline-\frac{1}{2}\log
		\left(\frac{\pi e}{2}\right),
\end{align*}
which gives the lower bound in~\eqref{eq:causalRx_bounds}.

Applying similar arguments for $\CnoncausalTxRx$ will obtain the LHS of~\eqref{eq:noncausal_bounds}.
This is shown in Appendix~\ref{sec:noncausal_bounds_proof}.

\subsubsection{Energy Arrival Information at the Transmitter Only}
We continue to the derivation of the lower bound on $\CcausalTx$, namely~\eqref{eq:causal_lower_bound}.
Fix $g^n\in\Gonline(b)$.
We construct an input distribution $P_{U^n}\in\mathcal{P}_n(b)$ that consists of independent strategy letters:
\[
P_{U^n}(u^n)=\prod_{t=1}^{n}P_{U_t}(u_t).
\]
Recall that each strategy letter is a function $U_t:\mathcal{E}^t\to\mathcal{X}$.
Therefore, it can also be viewed as a vector in $\mathcal{X}^{|\mathcal{E}|^t}$.
%=\mathbb{R}^{|\mathcal{E}|^t}$.
For each element of the vector, corresponding to each realization of $E^t$, we will choose the same distribution as in the previous case, namely:
\[
U_t(e^t)\sim P_X^{[g_t(e^t)]}.
\]
This will induce the same conditional distribution $P_{X^n\|E^n}$ on $X^n=U^n(E^n)$ that was constructed previously.
Additionally, since $g^n\in\Gonline(b)$, this input distribution is admissible, i.e. $P_{U^n}\in\mathcal{P}_n(b)$.

We can lower bound $I(U^n;Y^n)$ as follows:
\begin{align}
I(U^n;Y^n)
&=I(U^n;Y^n,E^n)-I(U^n;E^n|Y^n)\nonumber\\*
&\overset{\text{(i)}}{=}I(U^n;Y^n|E^n)-I(U^n;E^n|Y^n)\nonumber\\
&\overset{\text{(ii)}}{\geq} I(X^n;Y^n|E^n)-H(E^n),\label{eq:entropyE}
\end{align}
where (i) is because $U^n$ is independent of $E^n$; and (ii) is because $X^n=U^n(E^n)$ and the Markov chain $U^n-(E^n,X^n)-Y^n$.
Since $X^n$ is distributed according to the same $P_{X^n\|E^n}$ as before, we get
\[
\frac{1}{n}I(U^n;Y^n)\geq\T(g^n)-\frac{1}{2}\log\left(\frac{\pi e}{2}\right)-\frac{1}{n}H(E^n),
\]
which, after maximizing over $g^n\in\Gonline(b)$ and taking ${n\to\infty}$, gives:
\[
\CcausalTx\geq\Conline-\frac{1}{2}\log\left(\frac{\pi e}{2}\right)-H(E_t).
\]

It turns out, however, that this bound may be too loose.
The term $H(E_t)$ may become very large for different distributions of $E_t$, and is in fact unbounded for increasingly large alphabets~$\mathcal{E}$.
Intuitively, this gap implies that the receiver must learn the entire sequence of energy arrivals $E^n$ in order to know the codebook from which the transmitter chose the codeword. For example, the rate corresponding to \eqref{eq:entropyE} can be achieved by communicating the sequence of realizations of $E^n$ at the end of each block which will induce a rate penalty equal to the entropy rate of this process. However, this requirement can be made less strict by observing that our desired input distribution at each time depends only on~$g_t(e^t)$ - a deterministic function of $e^t$. By introducing special structure into $g^n$, we can make its entropy per symbol $\tfrac{1}{n}H(g^n(E^n))$, which is the amount of information that needs to be sent to the receiver, much smaller than $H(E_t)$. See Section~\ref{sec:entropyrate}.

Fix an online power control policy $g^n\in\Gonline(b)$.
We wish to construct $P_{U_t}$ in such a manner that $g_t(e^t)$ alone will determine $X_t$. %, i.e. that $g_t(E^t)$ will be a sufficient statistic to $X_t$.
This implies that for two different energy arrival realizations, say $e^t$ and $\check{e}^t$, that satisfy $g_t(e^t)=g_t(\check{e}^t)$, we wish to have $U_t(e^t)=U_t(\check{e}^t)$ with probability 1.
%Moreover, we will choose the same distribution as in the previous case, namely:
%\[
%U_t(e^t)\sim P_X^{[g_t(e^t)]}.
%\]

Since $U_t$ can be thought of as a vector of size $|\mathcal{E}|^t$,
we wish to specify the joint distribution of this multivariate random variable.
For that matter, define the set of all possible outcomes of the power control policy at time $t$:
\[
\mathcal{G}_t=\{g\in\mathbb{R}_+|\ g=g_t(e^t),\ e^t\in\mathcal{E}^t\}.
\]
This set defines a partition on the set $\mathcal{E}^t$, in the sense that disjoint subsets of $\mathcal{E}^t$ map to different $g\in\mathcal{G}_t$.
More precisely, let
\[
\mathcal{A}_t(g)=\{e^t\in\mathcal{E}^t|\ g_t(e^t)=g\}.
\]
Then $\mathcal{A}_t(g)$ for different $g$'s are disjoint and $\mathcal{E}^t=\cup_{g\in\mathcal{G}_t}\mathcal{A}_t(g)$.

%$\mathcal{A}_t(g)$ defines a subset of elements in the vector $U_t$, namely $\{U_t(e^t),\ e^t\in\mathcal{A}_t(g)\}$.
We will construct $P_{U_t}$ so that all the elements in each of these subsets will be equal with probability 1, and independent of all other elements of $U_t$:
For any $g\in\mathcal{G}_t$, let $Z_g$ be a random variable such that
\[
U_t(e^t)=Z_g\quad,\forall e^t\in\mathcal{A}_t(g)
\text{ w.p. 1},
\]
\[
Z_g\sim P_X^{[g]},
\]
and $Z_g$'s are independent for different $g$'s.

%We can now define $P_{U_t}$ according to the above discussion:
%\[
%P_{U_t}(u_t)=\prod_{g\in\mathcal{G}_t}
%P_X^{[g]}(u_t(g))
%\prod_{e^t:\ g_t(e^t)=g}
%1(u_t(e^t)=u_t(g)).
%\]
Note that, by construction, knowledge of $U_t$ and $g_t(E^t)$ suffices to know $X_t$:
\[
X_t=U_t(E^t)=U_t(g_t(E^t)).
\]
Clearly, $P_{U^n}\in\mathcal{P}_n(b)$ since $g^n\in\Gonline(b)$.

We proceed to lower bound $I(U^n;Y^n)$ for this distribution:
\begin{align}
	I(U^n;Y^n)
	&=
		I(U^n;Y^n,g^n(E^n))-I(U^n;g^n(E^n)|Y^n)\nonumber\\
	&\stackrel{\text{(i)}}{=}
		I(U^n;Y^n|g^n(E^n))-I(U^n;g^n(E^n)|Y^n)\nonumber\\
	&\geq
		I(U^n;Y^n|g^n(E^n))-H(g^n(E^n))\nonumber\\
	&\stackrel{\text{(ii)}}{=}
		I(X^n;Y^n|g^n(E^n))-H(g^n(E^n))\nonumber\\
	&\stackrel{\text{(iii)}}{\geq}
		I(X^n;Y^n|E^n)-H(g^n(E^n))
		\label{eq:causalTx_lower_bound_smith}
\end{align}
where (i) is because $U^n$ is independent of $g^n(E^n)$; (ii) is because $X^n=U^n(g^n(E^n))$ and the Markov chain $U^n-(g^n(E^n),X^n)-Y^n$; 
and (iii) is due to the Markov chain $E^n-(g^n(E^n),X^n)-Y^n$ and because $g^n(E^n)$ is a deterministic function of $E^n$.

Now, observe that our distribution $P_{U^n}$ on $U^n$ induces a distribution $P_{X^n\|E^n}$ on $X^n$ which is identical to the one we constructed in the previous case:
the joint distribution $P_{X^n,E^n,U^n}$ can be factored as 
\begin{align*}
P_{X^n,E^n,U^n}(x^n,e^n,u^n)
=\prod_{t=1}^{n}P_E(e_t)P_{U_t}(u_t)1\{x_t=u_t(e^t)\}.
\end{align*}
Summing over $u_n$:
\begin{align*}
\lefteqn{P_{X^n,E^n,U^{n-1}}(x^n,e^n,u^{n-1})}\\*
&=P_{X^{n-1},E^{n-1},U^{n-1}}(x^{n-1},e^{n-1},u^{n-1})
	\\* &\qquad\times
	\sum_{u_n}P_E(e_n)P_{U_n}(u_n)1\{x_n=u_n(e^n)\}\\
&=P_{X^{n-1},E^{n-1},U^{n-1}}(x^{n-1},e^{n-1},u^{n-1})
	\\* &\qquad\times
	\sum_{u_n(e^n)}P_E(e_n)P_{U_n(e^n)}(u_n(e^n))
	1\{x_n=u_n(e^n)\}\\
&=P_{X^{n-1},E^{n-1},U^{n-1}}(x^{n-1},e^{n-1},u^{n-1})
	P_E(e_n)P_X^{[g_n(e^n)]}(x_n).
\end{align*}
Summing over $u_{n-1}$, then $u_{n-2}$, and so forth, we obtain:
\[P_{X^n,E^n}(x^n,e^n)=\prod_{t=1}^{n}P_E(e_t)P_X^{[g_t(e^t)]}(x_t).\]

We can therefore apply~\eqref{eq:causalTxRx_online_lower_bound} to obtain
\begin{align*}
\frac{1}{n}I(U^n;Y^n)
\geq \T(g^n)
-\frac{1}{2}\log\left(\frac{\pi e}{2}\right)
-\frac{1}{n}H(g^n(E^n)).
\end{align*}
Finally, since $g^n$ was arbitrary we can maximize over all possible online policies:
\begin{align*}
\frac{1}{n}I(U^n;Y^n)&\geq\max_{g^n\in\Gonline(b)}
\{\T(g^n)-\tfrac{1}{n}H(g^n(E^n))\}
\\*&\hspace{12em}
-\frac{1}{2}\log\left(\frac{\pi e}{2}\right).
\end{align*}
Taking $n\to\infty$ and substituting in~\eqref{eq:causal_capacity}, we get~\eqref{eq:causal_lower_bound}.

%---------------------------%
\section{Conclusion}
%---------------------------%
\label{sec:conclusion}

We studied the communication problem with an energy harvesting transmitter over the AWGN channel. We characterized the information-theoretic capacity of this channel as an n-letter mutual information rate under various assumptions on the availability of energy arrival information. We  also considered the power control problem for energy harvesting communication that has been of interest in the recent communication theory literature and provided an approximately optimal solution for the online version of this problem. We then proceeded to connecting these two different formulations of the problem and showed that the information-theoretic capacity can be lower and upper bounded by the long-term average throughput, i.e. the solution of the power control problem. Putting these results together allowed us to approximate the information-theoretic capacity of the energy harvesting channel with a simple and insightful formula within a constant gap independent of system parameters.
 
There are many interesting research directions one can pursue from here. One immediate question is whether the approximation results in this paper can be significantly tightened to obtain better approximations for the capacity. One can also seek purely multiplicative approximations instead of the additive approximations we derived in this paper. Another interesting direction is to develop similar insights and results for energy harvesting processes with memory or certain correlation structure over time. Finally, the approximation approach developed in this paper can be used to understand the information-theoretic capacity as well as optimal online power control for various multi-user settings.

% use section* for acknowledgement
%\section*{Acknowledgment}

% trigger a \newpage just before the given reference
% number - used to balance the columns on the last page
% adjust value as needed - may need to be readjusted if
% the document is modified later
%\IEEEtriggeratref{8}
% The "triggered" command can be changed if desired:
%\IEEEtriggercmd{\enlargethispage{-5in}}

% references section

% can use a bibliography generated by BibTeX as a .bbl file
% BibTeX documentation can be easily obtained at:
% http://www.ctan.org/tex-archive/biblio/bibtex/contrib/doc/
% The IEEEtran BibTeX style support page is at:
% http://www.michaelshell.org/tex/ieeetran/bibtex/
%\bibliographystyle{IEEEtran}
% argument is your BibTeX string definitions and bibliography database(s)
%\bibliography{IEEEabrv,../bib/paper}
%
% <OR> manually copy in the resultant .bbl file
% set second argument of \begin to the number of references
% (used to reserve space for the reference number labels box)

\appendices

%---------------------------%
\section{Capacity and Maximum Throughput Do Not Depend on Initial Battery State}
%---------------------------%
\label{sec:initial_battery_state}

\subsection{Information-Theoretic Capacity Does Not Depend on Initial Battery State}
\label{subsec:capacity_initial_battery_state}
We prove Proposition~\ref{prop:initial_battery_state}, namely that the capacity does not depend on the initial state of the battery $B_0$.
Let $C$ be the capacity of the energy harvesting channel when $B_0$ is some arbitrary value in $[0,\bar{B}]$ unknown to the receiver, and let $C_{\bar{B}}$ be the capacity when $B_0=\bar{B}$.
We show that $C=C_{\bar{B}}$.

It is immediate that $C_{\bar{B}}\geq C$, since any achievable scheme for any value of $B_0$ can be achieved when $B_0=\bar{B}$ by ignoring the remaining energy in the battery.
To show $C_{\bar{B}}\leq C$, we show that any achievable scheme designed for $B_0=\bar{B}$ can be achieved in a system with arbitrary $B_0$.

We do so by transmitting a large number of zeros, therefore recharging the battery to $\bar{B}$, followed by the scheme designed for $B_0=\bar{B}$.
More precisely, we transmit $\ell$ zeros followed by an $(M,n,\varepsilon/2)$ code for $B_0=\bar{B}$.
Denote by $\mathcal{E}_1$ the event that the battery is not charged to $\bar{B}$, and by $\mathcal{E}_2$ the event that the code for $B_0=\bar{B}$ will produce an error.

First, we have
\begin{align*}
	\Pr\{\mathcal{E}_1\}
	&=\Pr\{B_0+\sum_{t=1}^{\ell}E_t<\bar{B}\}\\
	&\leq\Pr\{\sum_{t=1}^{\ell}E_t<\bar{B}\}\\
	&\leq\epsilon_\ell,
\end{align*}
where $\lim_{\ell\to\infty}\epsilon_\ell=0$, and the last inequality follows from the law of large numbers, using the fact that $\mathbb{E}[E_t]>0$.
We can choose $\ell$ large enough so that $\Pr\{\mathcal{E}_1\}\leq\varepsilon/2$. Note that this value of $\ell$ depends solely on $\bar{B}$ and the statistics of $E_t$, and does not depend on the actual value of $B_0$.

Next, from the i.i.d. nature of $E_t$, we have $\Pr\{\mathcal{E}_2|\mathcal{E}_1^c\}\leq\varepsilon/2$.
Therefore, the total probability of error for our scheme $\Pr\{\mathcal{E}_1\cup\mathcal{E}_2\}$ does not exceed $\varepsilon$.
The transmission spans $\ell+n$ channel uses, thus we have constructed an $(M,n+\ell,\varepsilon)$ code for the channel with arbitrary $B_0$.
By taking sufficiently large $n$, we can get a rate as close to $C_{\bar{B}}$ as desired.\qed

\subsection{Throughput Does Not Depend on Initial Battery State}
\label{subsec:power_allocation_initial_battery_state}

We state and prove the following proposition:
\begin{proposition}
\label{prop:power_allocation_initial_battery_state}
The long-term average throughput does not depend on the initial battery state, i.e. for any $b_1,b_2\in[0,\bar{B}]$:
\[
\liminf_{n\to\infty}\max_{g^n\in\mathcal{G}_n(b_1)}\T(g^n)
=\liminf_{n\to\infty}\max_{g^n\in\mathcal{G}_n(b_2)}\T(g^n),
\]
for offline and online policies alike.
\end{proposition}
This immediately implies that we can compute the throughput $\Conline$ or $\Coffline$ for any initial battery level, say $\bar{B}$, regardless of the actual battery level of interest $b_0$.
\begin{proof}
We will give the proof for online policies, however it transfers immediately to offline policies.
Denote 
\[\Conline(b)=\liminf_{n\to\infty}\max_{g^n\in\Gonline(b)}\T(g^n).\]
We will show that $\Conline(b)=\Conline(\bar{B})$ for any $0\leq b\leq\bar{B}$, which will imply the desired result.

First, clearly $\Conline(b)\leq\Conline(\bar{B})$ since $\Gonline(b)\subseteq\Gonline(\bar{B})$ for $0\leq b\leq\bar{B}$.
To show the other direction, let $\{\hat{g}^n\}_{n=1}^{\infty}$ be the sequence of maximal policies in $\Conline(\bar{B})$, that is:
\[
\hat{g}^n=\argmax_{g^n\in\mathcal{G}_n(\bar{B})}\T(g^n)
\qquad,n=1,2,\ldots
\]
Fix $\ell\geq1$. For any $n>\ell$, consider the following online power control policy $g^n$ for initial battery level $b$: Transmit zeros ($g_t=0$) for the first $\ell$ time slots. This will allow the battery to completely recharge to $\bar{B}$ with high probability. Then, if $B_\ell=\bar{B}$, transmit the policy $\hat{g}^{n-\ell}$. Otherwise, transmit zeros for $n-\ell$ time slots (i.e. give up on the entire transmission).
More precisely, define the new policy as follows, for $t=1,\ldots,n$:
\[
g_t(e^t)=\begin{cases}
0&,1\leq t\leq \ell\\
1_{\{B_\ell=\bar{B}\}}\cdot \hat{g}_{t-\ell}(e_{\ell+1}^{t})
&,\ell+1\leq t
\end{cases}
\]
where $1_{\{\cdot\}}$ is the indicator function.
Observe that $B_\ell$ is a deterministic function of $e^\ell$, which is given by
$B_\ell=\min\big\{b+\sum_{t=1}^{\ell}e_t,\bar{B}\big\}$.
We have for any $n>\ell$:
\begin{align*}
\T(g^n)
&=\frac{1}{n}\sum_{t=1}^{n}\mathbb{E}\left[\frac{1}{2}\log\big(1+g_t(E^t)\big)\right]\\
&=\frac{1}{n}\sum_{t=\ell+1}^{n}\mathbb{E}\left[\frac{1}{2}\log\big(1+1_{\{B_\ell=\bar{B}\}}\cdot\hat{g}_{t-\ell}(E_{\ell+1}^t)\big)\right]\\
&=\frac{1}{n}\sum_{t=1}^{n-\ell}\mathbb{E}\left[1_{\{B_\ell=\bar{B}\}}\cdot\frac{1}{2}\log\big(1+\hat{g}_t(E_{\ell+1}^{\ell+t})\big)\right]\\
&\overset{\text{(i)}}{=}\frac{1}{n}\sum_{t=1}^{n-\ell}\Pr\{B_\ell=\bar{B}\}\cdot\mathbb{E}\left[\frac{1}{2}\log(1+\hat{g}_t(E_{\ell+1}^{\ell+t})\big)\right]\\
&\overset{\text{(ii)}}{=}\Pr\{B_\ell=\bar{B}\}\cdot\frac{1}{n}\sum_{t=1}^{n-\ell}\mathbb{E}\left[\frac{1}{2}\log(1+\hat{g}_t(E^{t})\big)\right]\\
&=\Pr\{B_\ell=\bar{B}\}\cdot\frac{n-\ell}{n}\T(\hat{g}^{n-\ell}),
\end{align*}
where (i) is because $B_\ell$ depends only on $E^\ell$, and $E_t$ is independent over time;
and (ii) is because $E_t$ is i.i.d.

Note that $g^n\in\Gonline(b)$ for any $0\leq b\leq\bar{B}$. Therefore:
\begin{align}
\Conline(b)
&\geq \liminf_{n\to\infty}\T(g^n)\nonumber\\
&=\Pr\{B_\ell=\bar{B}\}\cdot\liminf_{n\to\infty}\frac{n-\ell}{n}\T(\hat{g}^{n-\ell})\nonumber\\
&=\Pr\{B_\ell=\bar{B}\}\cdot\Conline(\bar{B}).
\label{eq:Tb_geq_plTBbar}
\end{align}
We can lower-bound the probability of recharging the battery using Chebyshev's inequality:
\begin{align*}
\Pr\{B_\ell=\bar{B}\}
&=1-\Pr\big\{b+\sum_{t=1}^{\ell}E_t<\bar{B}\big\}\\
&\geq 1 - \frac{\ell\cdot\text{Var}(E_t)}{(\ell\cdot\mathbb{E}[E_t]-\bar{B}+b)^2}\\
&\triangleq 1-\epsilon_\ell,
\end{align*}
where $\epsilon_\ell\to0$ as $\ell\to\infty$.
Substituting this in~\eqref{eq:Tb_geq_plTBbar} yields $\Conline(b)\geq(1-\epsilon_\ell)\cdot\Conline(\bar{B})$.
Since the LHS does not depend on $\ell$, we can take
$\ell\to\infty$ to obtain $\Conline(b)\geq\Conline(\bar{B})$, which concludes the proof.
\end{proof}

%---------------------------%
\section{Capacity with Noncausal Side Information}
%---------------------------%
\label{sec:capacity_noncausal_Rx}

We prove Theorem~\ref{thm:capacity} for the case of energy arrival information available noncausally at the receiver and the transmitter~\eqref{eq:noncausal_capacity}. Recall the definition of $\mathcal{F}_n(b)$ in~\eqref{eq:noncausal_Fn_definition} and
fix $P_{X^n|E^n}\in\mathcal{F}_n(b)$.
We transmit $k$ blocks of length $n+\ell+1$ as in Section~\ref{sec:capacity_proof}.
Generate a random codebook for every $e^{nk}\in\mathcal{E}^{nk}$ by generating $k$ independently drawn codewords from $P_{X^n|E^n}$.
Specifically, denoting $\mathbf{e}_i=e_{(i-1)(n+\ell+1)+1}^{(i-1)(n+\ell+1)+n}$, we choose $\mathbf{x}_i(\mathbf{e}_i)\sim P_{X^n|E^n}(\cdot|\mathbf{e}_i)$ and transmit
\[
	x_{(i-1)(n+\ell+1)+1}^{i(n+\ell+1)}=
	[z_i\cdot\mathbf{x}_i(\mathbf{e}_i),\ 
	\sqrt{b_{(i-1)(n+\ell+1)+n+1}},\ \mathbf{0}],
\]
where $z_i=1$ if $b_{(i-1)(n+\ell+1)}\geq b$ and $z_i=0$ otherwise, and $\mathbf{0}$ is a length-$\ell$ vector of zeros.
Similarly to Section~\ref{sec:capacity_proof}, the energy constraint will be satisfied.
From here on, we repeat the arguments of Section~\ref{sec:capacity_proof}.
The receiver makes use of $\mathbf{y}_i=y_{(i-1)(n+\ell+1)+1}^{(i-1)(n+\ell+1)+n}$ and $\mathbf{e}_i$, $i=1,\ldots,k$, for decoding.
The channel is memoryless with i.i.d. side information available at both the receiver and the transmitter.
Note that the i.i.d. Bernoulli RV $Z_i$ is independent of the side information $\mathbf{e}_i$.
Therefore we obtain
\begin{align}
	\CnoncausalTxRx\geq\limsup_{n\to\infty}\frac{1}{n}
		\sup_{P_{X^n|E^n}\in\mathcal{F}_n(b)}
		I(X^n;Y^n|E^n).
		\label{eq:noncausal_achievability}
\end{align}

Conversely, from Fano's inequality we have
\[
	\CnoncausalTxRx\leq\liminf_{n\to\infty}
	\frac{1}{n}\sup_{P_{X^n|E^n}\in\mathcal{F}_n(b)}
	I(X^n;Y^n|E^n),
\]
which, combined with~\eqref{eq:noncausal_achievability}, gives \eqref{eq:noncausal_capacity}.

%---------------------------%
\section{Alternative Representation of Energy Constraints}
%---------------------------%
\label{sec:alternative_constraints}

In this section, we derive the alternative representation of the energy constraints stated in the beginning of Section~\ref{subsec:upper_bounds}.
Suppose $x^n$ and $e^n$ satisfy constraints~\eqref{eq:EH_constraint} and~\eqref{eq:EH_battery} for $t=1,\ldots,n$ and $B_0=b$, that is
\begin{align}
x_t^2&\leq b_t,\label{eq:battery_constraint_lowercase}\\
b_t&=\min\{b_{t-1}-x_{t-1}^2+e_t,\bar{B}\},\label{eq:battery_evolution_lowercase}
\end{align}
for $t=1,\ldots,n$, where $b_0=b$.
We show that this is equivalent to satisfying
\begin{align}
	\sum_{j=i}^{t}x_j^2&\leq\bar{B}+\sum_{j=i+1}^{t}e_j
	&,i=1,\ldots,t\label{eq:sum_Bbar}\\
	\sum_{j=1}^{t}x_j^2&\leq b+\sum_{j=1}^{t}e_j
	\label{eq:sum_total_energy}
\end{align}
for $t=1,\ldots,n$.

Suppose $x^n,e^n$ satisfy \eqref{eq:battery_constraint_lowercase} and \eqref{eq:battery_evolution_lowercase}.
For any $t=1,\ldots,n$:
\begin{align}
	x_t^2&\leq\bar{B},\label{eq:t_Bbar}\\
	x_t^2&\leq b_{t-1}-x_{t-1}^2+e_t,\label{eq:t_bt}
\end{align}
where~\eqref{eq:t_Bbar} gives~\eqref{eq:sum_Bbar} for $i=t$, and~\eqref{eq:t_bt} can be further written as
\begin{align}
	x_{t-1}^2+x_{t}^2&\leq\bar{B}+e_t,\label{eq:t-1_Bbar}\\
	x_{t-1}^2+x_{t}^2&\leq b_{t-2}-x_{t-2}^2+e_{t-1}+e_t.
	\label{eq:t-1_bt}
\end{align}
\eqref{eq:t-1_Bbar} gives~\eqref{eq:sum_Bbar} for $i=t-1$, and~\eqref{eq:t-1_bt} can be written as
\[
	x_{t-2}^2+x_{t-1}^2+x_t^2\leq \min\{b_{t-3}-
	x_{t-3}^2+e_{t-2},\bar{B}\}+e_{t-1}+e_t.
\]
Continuing in this fashion gives~\eqref{eq:sum_Bbar} for all $i\leq t$ and~\eqref{eq:sum_total_energy}.

Now, let $x^n,e^n$ satisfy \eqref{eq:sum_Bbar} and \eqref{eq:sum_total_energy} for $t=1,\ldots,n$.
Applying \eqref{eq:battery_evolution_lowercase} for $b_1,\ldots,b_t$, we can express the battery state at time $t$ as follows:
\[
b_t=\min\Big\{b-\sum_{i=1}^{t-1}x_i^2+\sum_{i=1}^{t}e_i,\ 
\min_{1\leq i\leq t}\big\{\bar{B}-\sum_{j=i}^{t-1}x_j^2+\sum_{j=i+1}^{t}e_j\big\}\Big\}.
\]
Hence, \eqref{eq:battery_constraint_lowercase} holds if and only if:
\begin{align*}
x_t^2&\leq\bar{B}-\sum_{j=i}^{t-1}x_j^2+\sum_{j=i+1}^{t}e_j
&,i=1,\ldots,t,\\
x_t^2&\leq b-\sum_{i=1}^{t-1}x_i^2+\sum_{i=1}^{t}e_i,
\end{align*}
for $t=1,\ldots,n$, which is exactly \eqref{eq:sum_Bbar} and \eqref{eq:sum_total_energy}.

%---------------------------%
\section{Noncausal Capacity Bounds}
%---------------------------%
\label{sec:noncausal_bounds_proof}

We derive the bounds on the capacity with noncausal observations of the energy arrivals at the transmitter and the receiver, namely~\eqref{eq:noncausal_bounds}.
We repeat the steps of Section~\ref{sec:capacity_bounds}, starting with the upper bound.
Rewrite~\eqref{eq:noncausal_Fn_definition} as (see Appendix~\ref{sec:alternative_constraints}):
\begin{align*}
	\mathcal{F}_n(b)=\Big\{&
	P_{X^n|E^n}\text{ s.t. $\forall e^n\in\mathcal{E}^n$, 
	a.s. for $t=1,\ldots,n$:}\\*
	&\sum_{j=i}^{t}X_j^2\leq\bar{B}+\sum_{j=i+1}^{t}e_j
	\quad,i=1,\ldots,t,\nonumber\\*
	&\sum_{j=1}^{t}X_j^2\leq b+\sum_{j=1}^{t}e_j
	\Big\}.
\end{align*}
%Note that as in Appendix~\ref{subsec:upper_bounds}, $\mathcal{F}_n(b)$ (cf.~\eqref{eq:noncausal_Fn_definition}) can be written as the set of all $P_{X^n|E^n}$ such that $\forall e^n\in\mathcal{E}^n$, a.s. for every $t=1,\ldots,n$:
%\begin{align}
%	\sum_{j=i}^{t}X_j^2&\leq\bar{B}+\sum_{j=i+1}^{t}e_j
%	&,i=1,\ldots,t\nonumber\\*
%	\sum_{j=1}^{t}X_j^2&\leq b+\sum_{j=1}^{t}e_j.
%\end{align}
%For details of the derivation see Appendix~\ref{sec:alternative_constraints}.
Additionally, define 
\begin{align*}
	\mathcal{F}_n^\ast(b)=\Big\{&P_{X^n|E^n}
	\text{ s.t. for $t=1,\ldots,n$ and
	$\forall e^n\in\mathcal{E}^n$:}\\*
	&\sum_{j=i}^{t}\mathbb{E}[X_j^2|E^n=e^n]\leq\bar{B}+\sum_{j=i+1}^{t}e_j
	%\\*&\hspace{12em}
	\ 
	,i=1,\ldots,t,\\*
	&\sum_{j=1}^{t}\mathbb{E}[X_j^2|E^n=e^n]\leq b+\sum_{j=1}^{t}e_j\Big\}.
\end{align*}
Observe that ${\mathcal{F}}_n(b)\subseteq{\mathcal{F}}_n^\ast(b)$.

Similarly, we can write~\eqref{eq:G_definition} for the offline policies as
\begin{align*}
	\Goffline(b)
	=\Big\{&g^n=(g_1,\ldots,g_n),\ 
	g_t:\mathcal{E}^n\to\mathbb{R}_+,\\*
	&\text{s.t. }\forall e^n\in\mathcal{E}^n:\\*
	&\sum_{j=i}^{t}g_j(e^n)\leq\bar{B}+\sum_{j=i+1}^{t}e_j
		\quad,i=1,\ldots,t,\\*
	&\sum_{j=1}^{t}g_j(e^n)\leq b+\sum_{j=1}^{t}e_j
		\quad,t=1,\ldots,n\Big\}.
\end{align*}

We now upper bound the expression in~\eqref{eq:noncausal_capacity}, repeating the steps in Section~\ref{subsec:upper_bounds}.
First we upper bound the mutual information as:
\[
I(X^n;Y^n|E^n)\leq\sum_{t=1}^{n}I(X_t;Y_t|E^n).
\]
Next, we apply this inequality to~\eqref{eq:noncausal_capacity}:
\begin{align*}
\CnoncausalTxRx
	&=
		\lim_{n\to\infty}\frac{1}{n}
		\sup_{P_{X^n|E^n}\in\mathcal{F}_n(b)}
		I(X^n;Y^n|E^n)\\
	&\leq
		\liminf_{n\to\infty}\frac{1}{n}
		\sup_{P_{X^n|E^n}\in\mathcal{F}_n(b)}
		\sum_{t=1}^{n}I(X_t;Y_t|E^n)\\
	&\leq
		\liminf_{n\to\infty}\frac{1}{n}
		\sup_{P_{X^n|E^n}\in\mathcal{F}^*_n(b)}
		\sum_{t=1}^{n}I(X_t;Y_t|E^n).
\end{align*}
For each $e^n\in\mathcal{E}^n$, we have:
\[
I(X_t;Y_t|E^n=e^n)
\leq \frac{1}{2}\log\big(1+\mathbb{E}[X_t^2|E^n=e^n]\big),
\]
and thus
\begin{align*}
\CnoncausalTxRx
	&\leq
		\liminf_{n\to\infty}\frac{1}{n}
		\sup_{P_{X^n|E^n}\in\mathcal{F}_n^\ast(b)}
		\mathbb{E}\left[\sum_{t=1}^{n}
		\frac{1}{2}\log(1+\mathbb{E}[X_t^2|E^n])
		\right]\\
	&\leq
		\liminf_{n\to\infty}\frac{1}{n}
		\max_{g^n\in\Goffline(b)}
		\mathbb{E}\left[\sum_{t=1}^{n}
		\frac{1}{2}\log(1+g_t(E^n))\right]\\
	&=
		\Coffline.
\end{align*}

For the lower bound,
fix $g^n\in\Goffline(b)$.
Let 
\[
P_{X^n|E^n}(x^n|e^n)=\prod_{t=1}^{n}P_{X_t|E^n}(x_t|e^n),
\]
where $P_{X_t|E^n}(x_t|e^n)=P^{[g_t(e^n)]}_{X}(x_t)$ and $P^{[S]}_X$ is a distribution with support $[-\sqrt{S},\sqrt{S}]$.
Clearly $P_{X^n|E^n}\in\mathcal{F}_n(b)$, and
\[
I(X^n;Y^n|E^n)=\sum_{t=1}^{n}I(X_t;Y_t|E^n).
\]
Then, after maximizing over all such distributions, we get $I(X_t;Y_t|E^n=e^n)=\CSmith(g_t(e^n))$, and therefore:
\begin{align}
\CnoncausalTxRx
	&\geq
		\limsup_{n\to\infty}\frac{1}{n}
		\mathbb{E}\left[\sum_{t=1}^{n}
		\CSmith(g_t(e^n))\right]
		\label{eq:noncausal_lower_bound_smith}\\
	&\geq
		\liminf_{n\to\infty}\frac{1}{n}
		\mathbb{E}\left[\sum_{t=1}^{n}
		\frac{1}{2}\log(1+g_t(E^n))\right]
		-\frac{1}{2}\log\left(\frac{\pi e}{2}\right),
		\nonumber
\end{align}
where the second inequality is due to Lemma~\ref{lemma:EPI_lower_bound}.
%~\eqref{eq:ozarow_wyner_lower_bound}.
Since this is true for any ${g^n\in\Goffline(b)}$, we can take the maximum to obtain~\eqref{eq:noncausal_bounds}.

\section{Multiplicative Bounds: Proof of Theorem~\ref{thm:mult_bounds}}
%---------------------------%
\label{sec:mult_bounds}

Recall the proof of the lower bound in Section~\ref{subsec:lower_bounds}.
We will continue from equation~\eqref{eq:lower_bound_smith}.
We will develop here a different lower bound for 
$\CSmith(S)$, specifically a \emph{multiplicative} lower bound.
%\[
%	C_\text{Smith}(S)=
%	\max_{X^2\leq S}
%	I(X;X+N),
%\]
%where $N\sim\mathcal{N}(0,1)$.
In what follows, we will show that 
\[\CSmith(S)\geq\eta\cdot\frac{1}{2}\log(1+S),\]
for an appropriate $\eta$ and all $S\geq0$.
Substituting this in~\eqref{eq:lower_bound_smith}, \eqref{eq:causalTx_lower_bound_smith}, and~\eqref{eq:noncausal_lower_bound_smith}, immediately yields equations~\eqref{eq:mult_bound_causal}-\eqref{eq:mult_bound_noncausal}.

First, one can obtain from~\eqref{eq:EPI_lower_bound} in the proof of Lemma~\ref{lemma:EPI_lower_bound} the following lower bound:
\[
	\frac{C_{\text{Smith}}(S)}
		{\frac{1}{2}\log(1+S)}
	\geq\frac{\frac{1}{2}\log\left(1+
	\frac{2}{\pi e}S\right)}
		{\frac{1}{2}\log(1+S)}
	\geq\frac{2}{\pi e},
\]
which implies $\eta\geq\frac{2}{\pi e}=0.2342$.
However, it can be observed numerically that $\eta$ is larger that this value, and it is in fact $\eta=0.7473$.
In what follows, we show $\frac{C_\text{Smith}(S)}{\frac{1}{2}\log(1+S)}\geq\eta$ via a numerical proof.
We divide $\mathbb{R}_+$ into five regions, and show the inequality holds for all $S$ in each region.
%\begin{enumerate}

\subsection{\texorpdfstring{$0\leq S\leq 0.69$}{0<S<0.69}}
Consider a binary input distribution, that is $X=\pm\sqrt{S}$ w.p. $1/2$.
Denote 
\[C_{\text{bin}}(S)\triangleq
I(X;X+N)=I(Z;\sqrt{S}Z+N),\]
where $Z=\pm1$ w.p. $1/2$.
Then
\[
	\frac{\CSmith(S)}
		{\frac{1}{2}\log(1+S)}
	\geq\frac{C_{\text{bin}}(S)}
		{\frac{1}{2}\log(1+S)}
	\geq\frac{C_{\text{bin}}(S)}
		{\frac{1}{2\ln 2}S}.
\]
By~\cite[Lemma 1]{GuoShamaiVerdu2005}, we have
$C_{\text{bin}}(S)=\frac{S}{2\ln 2}+o(S)$, where $\frac{o(S)}{S}\to0$ as $S\to0$.
This implies
$R(S)\triangleq\frac{C_{\text{bin}}(S)}{\frac{1}{2\ln2}S}\to1$, and along with $\CSmith(S)\leq\frac{1}{2}\log(1+S)$ we conclude
\[
	\frac{\CSmith(S)}
		{\frac{1}{2}\log(1+S)}
	\to 1
	\quad\text{when}\quad
	S\to0.
\]
In fact, this was already observed by Shannon in his 1948 paper~\cite{Shannon1948}.

Now, we will show that $R(S)=\frac{C_{\text{bin}}(S)}{\frac{1}{2\ln2}S}$ is non-increasing.
By~\cite[Corollary 1]{GuoShamaiVerdu2005}, the function $C_{\text{bin}}(S)$ is concave, implying that the derivative $C'_{\text{bin}}(S)$ is non-increasing.
By the mean value theorem, for every $S>0$ there is some $0\leq c\leq S$ such that
\[
C'_{\text{bin}}(S)
\leq C'_{\text{bin}}(c)
=\frac{C_{\text{bin}}(S)-C_{\text{bin}}(0)}
{S-0}=\frac{C_{\text{bin}}(S)}{S}.
\]
Next, we take the derivative of $R(S)$:
\[
	R'(S)
	=\frac{2\ln2}{S}\left(
	C'_{\text{bin}}(S)
	-\frac{C_{\text{bin}}(S)}{S}
	\right)
	\leq 0,
\]
observing that $R(S)$ is monotonic non-increasing.

We compute $R(S)$ for $S=0.69$ numerically to obtain $R(0.69)=0.7501$.
Since $R(S)$ is non-increasing and $R(0)=1$, this implies
\[
\frac{\CSmith(S)}{\frac{1}{2}\log(1+S)}\geq R(S)>0.75
\]
for $0\leq S\leq0.69$.

\subsection{\texorpdfstring{$0.69\leq S\leq 170$}{0.69<S<170}}
In this part, we compute $\CSmith(S)$ for a finite set of points $\{S_i\}_{i=1}^{N}$ in $[0.5,170]$, where $0.5=S_1<S_2<\ldots<S_N=170$.
For every $S_i\leq S\leq S_{i+1}$, $i=1,\ldots,N-1$:
\[
	\frac{\CSmith(S)}
		{\frac{1}{2}\log(1+S)}
	\geq\frac{\CSmith(S_i)}
		{\frac{1}{2}\log(1+S_{i+1})}.
\]
Computing this lower bound for any such set of points and taking the minimal value will give a lower bound for all $S\in[0.5,170]$.
We compute this numerically using the algorithm suggested in~\cite{Smith1971} and obtain
\[
	\frac{\CSmith(S)}
		{\frac{1}{2}\log(1+S)}
	\geq 0.7473,
	\hspace{3em}
	0.5\leq S\leq 170.
\]
%The numerical results are plotted in Figure~\ref{fig:mult_bounds}.

\subsection{\texorpdfstring{$170\leq S\leq 195$}{170<S<195}}
For all $170\leq S\leq 195$, we have
\[
	\frac{\CSmith(S)}
		{\frac{1}{2}\log(1+S)}
	\geq\frac{\CSmith(170)}
		{\frac{1}{2}\log(1+S)}
	\geq\frac{\CSmith(170)}
		{\frac{1}{2}\log(1+195)}
	=0.7519.
\]
%The function $\CSmith(170)/\frac{1}{2}\log(1+S)$ is plotted for $170\leq S\leq195$ in Figure~\ref{fig:mult_bounds}.

\subsection{\texorpdfstring{$195\leq S\leq 340$}{195<S<340}}
Let $C_{\mathrm{unif}}(S)\triangleq I(X;X+N)$ where $X\sim U[-\sqrt{S},\sqrt{S}]$.
As before, we compute $C_{\mathrm{unif}}(S)$ numerically for a set of points $\{S_j\}_{j=1}^{M}$, where $195=S_1<\ldots<S_M=340$.
This involves numerical integration of the form $\int f(y)\log f(y)dy$, where $f(y)=
\frac{1}{2\sqrt{S}}(Q(y-\sqrt{S})
-Q(y+\sqrt{S}))$ and
$Q(x)\triangleq\frac{1}{\sqrt{2\pi}}\int_{x}^{\infty}e^{-u^2/2}du$.
For every $S_j\leq S\leq S_{j+1}$, $j=1,\ldots,M-1$:
\[
	\frac{\CSmith(S)}
		{\frac{1}{2}\log(1+S)}
	\geq \frac{C_{\mathrm{unif}}(S_j)}
		{\frac{1}{2}\log(1+S_{j+1})}.
\]
Computing this lower bound numerically for the set of points $S_j=195+(j-1)0.5$, $j=1,\ldots,291$, and taking the minimum, gives:
\[
	\frac{\CSmith(S)}
		{\frac{1}{2}\log(1+S)}
	\geq 0.7482,
	\hspace{3em}
	195\leq S\leq 340.
\]
%The numerical results are plotted in Figure~\ref{fig:mult_bounds}.

\subsection{\texorpdfstring{$340\leq S$}{340<S}}
For $S\geq340$, we use Lemma~\ref{lemma:EPI_lower_bound} to see that
\begin{align*}
	\frac{\CSmith(S)}
		{\frac{1}{2}\log(1+S)}
	&\geq\frac{\frac{1}{2}\log(1+S)
		-\frac{1}{2}\log\left(\frac{\pi e}{2}\right)}
		{\frac{1}{2}\log(1+S)}\\
	&\geq 1-\frac{\log\left(\frac{\pi e}{2}\right)}
		{\log(1+340)}\\
	&=0.7511.
\end{align*}

Combining all the above bounds, we conclude that
\[
	\frac{\CSmith(S)}
	{\frac{1}{2}\log(1+S)}
	\geq 0.7473,
\]
for all $S\geq 0$.\qed

\begin{figure}
\centering
\includegraphics[width=0.45\textwidth]{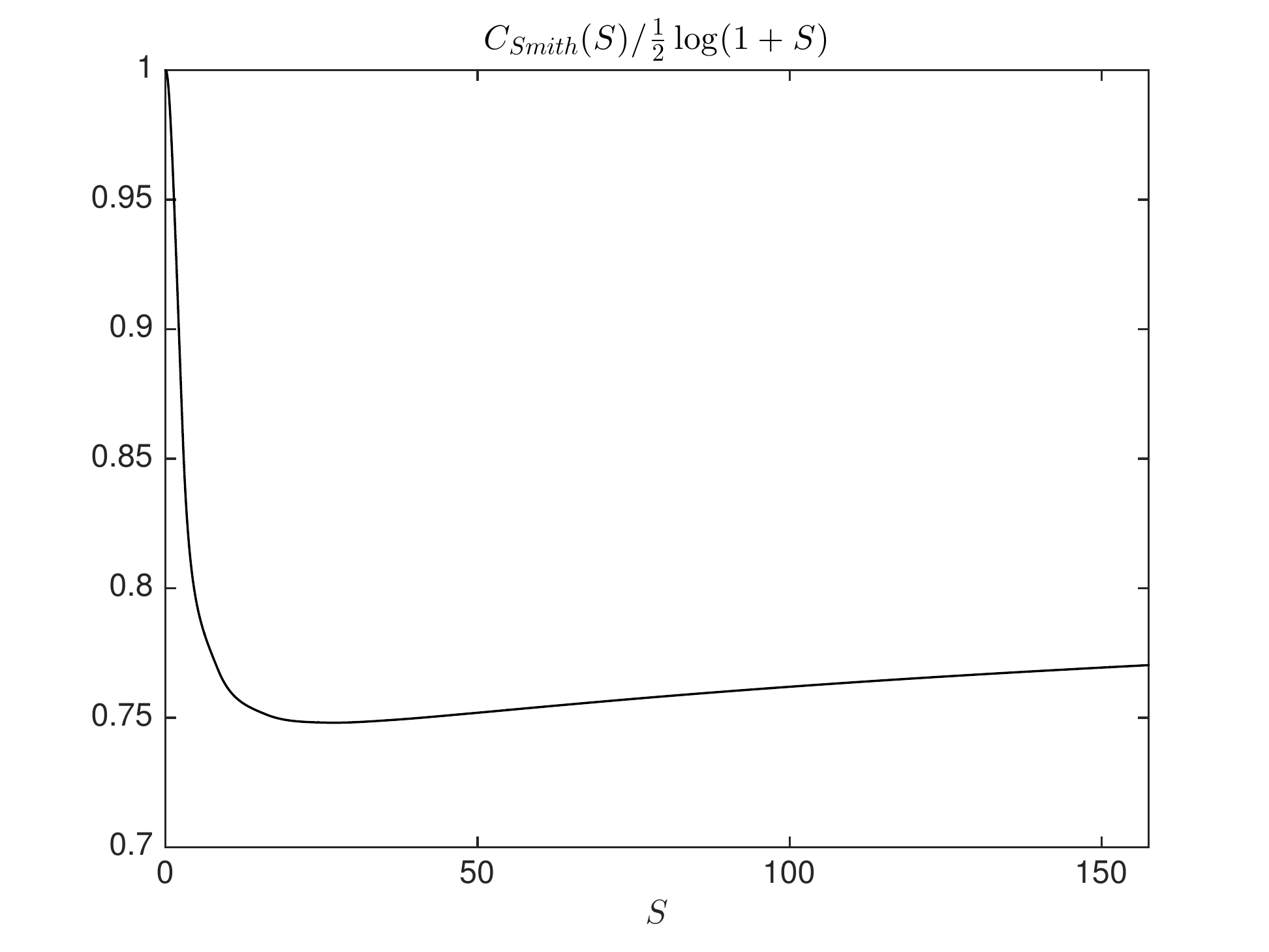}
%\caption{Lower bounds on the ratio $\CSmith(S)/\frac{1}{2}\log(1+S)$, for different regions of the parameter $S$. The minimal value in the graph is 0.7473.}
\caption{Numerical evaluation of the ratio $\CSmith(S)/\frac{1}{2}\log(1+S)$.}
\label{fig:mult_bounds}
\end{figure}

\section{Upper Bound on Expected Epoch Length in Generalized Bernoulli Policy}
\label{sec:L_finite_expectation}

As shown in Section~\ref{subsubsec:general_energy_arrivals}, the epoch length in the \emph{Generalized Bernoulli} policy is given by
\[
L=\{\inf\ t:\ S_t\geq\bar{B}[1-(1-q)^t]\},
\]
where $S_t=\sum_{i=2}^{t+1}\tilde{E}_i$ and $\tilde{E}_i=\min\{E_i,\bar{B}\}$.
Consider the following RV:
\[
\tilde{L}=\{\inf\ t:\ S_t\geq\bar{B}\}.
\]
Clearly $L\leq \tilde{L}$, therefore:
\begin{align}
\mathbb{E}L&\leq\mathbb{E}\tilde{L}\nonumber\\
&=\sum_{\ell=1}^{\infty}\Pr(\tilde{L}\geq\ell)\nonumber\\
&=\sum_{\ell=1}^{\infty}\Pr(S_1<\bar{B},S_2<\bar{B},\ldots,S_{\ell-1}<\bar{B})\nonumber\\
&=\sum_{\ell=1}^{\infty}\Pr(S_{\ell-1}<\bar{B}),\label{eq:ELtilde_sum}
\end{align}
where the last equality is due to the non-negativity of the energy arrivals $\tilde{E}_t\geq0$.
Using Chernoff bound, we upper bound the probability in the sum, for any $\theta>0$:
\begin{align*}
\Pr(S_{\ell-1}<\bar{B})&\leq \mathbb{E}[e^{\theta(\bar{B}-S_{\ell-1})}]\\
&=e^{\theta\bar{B}}\big(\mathbb{E}[e^{-\theta\tilde{E}_t}]\big)^{\ell-1}.
\end{align*}
Substituting the in~\eqref{eq:ELtilde_sum}, we have an infinite series of the form:
\[
\mathbb{E}L\leq \sum_{l=1}^{\infty}e^{\theta\bar{B}}\big(\mathbb{E}[e^{-\theta\tilde{E}_t}]\big)^{\ell-1}.
\]
Now, since $\tilde{E}_t\geq0$, we must have $e^{-\theta\tilde{E}_t}\leq1$ w.p.~1.
This implies that $\mathbb{E}[e^{-\theta\tilde{E}_t}]=1$ if and only if $\tilde{E}_t=0$ w.p.~1. However, since $\tilde{E}_t>0$ with positive probability, we must have $\mathbb{E}[e^{-\theta\tilde{E}_t}]<1$.
Hence the series must converge to a finite number and $\mathbb{E}L<\infty$.

\bibliographystyle{IEEEtran}

\bibliography{energy_harvesting}

\end{document}